\documentclass[prd,onecolumn,aps,nofootinbib,eqsecnum,notitlepage,nobibnotes,superscriptaddress,preprintnumbers,floatfix]{revtex4-1}

\usepackage[utf8]{inputenc}
\usepackage{color}
\usepackage{amsfonts}
\usepackage{graphicx}
\usepackage[dvipsnames]{xcolor}
\definecolor{refs}{RGB}{245,156,74}
 \usepackage[colorlinks=true,hyperfootnotes=true,citecolor=cyan]{hyperref}
\usepackage{tabularx,booktabs}
\newcolumntype{Y}{>{\centering\arraybackslash}X}

\usepackage{ulem}
\usepackage{dsfont} 

\pdfoutput=1

\usepackage{amsmath,amsthm,amssymb}
\usepackage{bm}

\usepackage{dcolumn}
\allowdisplaybreaks[1]

\begin{document}
\newcommand{\newc}{\newcommand}

\newc{\rk}[1]{{\color{red} #1}}
\newc{\ben}{\begin{eqnarray}}
\newc{\een}{\end{eqnarray}}
\newc{\be}{\begin{equation}}
\newc{\ee}{\end{equation}}
\newc{\ba}{\begin{eqnarray}}
\newc{\ea}{\end{eqnarray}}
\newc{\D}{\partial}
\newc{\rH}{{\rm H}}
\newc{\cH}{{\mathcal H}}
\newc{\dphi}{\delta\phi}
\newc{\pa}{\partial}
\newc{\tp}{\dot{\phi}}
\newc{\ttp}{\ddot{\phi}}
\newc{\drhoc}{\delta\rho_c}
\newc{\aB}{\alpha_{\rm B}}
\newc{\aK}{\alpha_{\rm K}}
\newc{\aM}{\alpha_{\rm M}}
\newc{\bn}{\beta_{n_c}}
\newc{\bK}{\beta_{\rm K}}
\newc{\delc}{\delta_{c{\rm N}}}
\newc{\eH}{\epsilon_{\rm H}}
\newc{\ep}{\epsilon_{\phi}}

\newc{\cs}{c_{\rm s}}
\newc{\dd}{{\rm d}}

\newcommand{\deltadm}{\delta_{\rm dm}}
\newcommand{\thetadm}{\theta_{\rm dm}}
\newcommand{\deltade}{\delta_{\rm de}}
\newcommand{\thetade}{\theta_{\rm de}}
\newcommand{\rhode}{\rho_{\rm de}}
\newcommand{\rhodm}{\rho_{\rm dm}}
\newcommand{\Omegadr}{\Omega_{\rm DR}}
\newcommand{\Hunits}{\,{\rm km\,s}^{-1}\,{\rm Mpc}^{-1}}

\title{Probing elastic interactions in the dark sector and the role of $S_8$}

\author{Jose Beltr\'an Jim\'enez}
\email{jose.beltran@usal.es}
\affiliation{Departamento~de~F{\'i}sica~Fundamental~and~IUFFyM,~Universidad~de~Salamanca,~E-37008~Salamanca,~Spain.}
\author{Dario Bettoni}
\email{bettoni@usal.es}
\affiliation{Departamento~de~F{\'i}sica~Fundamental~and~IUFFyM,~Universidad~de~Salamanca,~E-37008~Salamanca,~Spain.}
\author{David Figueruelo}
\email{davidfiguer@usal.es}
\affiliation{Departamento~de~F{\'i}sica~Fundamental~and~IUFFyM,~Universidad~de~Salamanca,~E-37008~Salamanca,~Spain.}
\author{Florencia A. Teppa Pannia}
\email{f.a.teppa.pannia@usal.es}
\affiliation{Departamento~de~F{\'i}sica~Fundamental~and~IUFFyM,~Universidad~de~Salamanca,~E-37008~Salamanca,~Spain.}
\author{Shinji Tsujikawa}
\email{tsujikawa@waseda.jp}
\affiliation{Department of Physics, Waseda University, 3-4-1 Okubo, Shinjuku, Tokyo 169-8555, Japan.}

\begin{abstract}
We place observational constraints on two models within a class of scenarios featuring an elastic interaction between dark energy and dark matter that only produces momentum exchange up to first order in cosmological perturbations. The first one corresponds to a perfect-fluid model of the dark components with an explicit interacting Lagrangian, where dark energy acts as a dark radiation at early times and behaves as a cosmological constant at late times. The second one is a dynamical dark energy model with a dark radiation component, where the momentum exchange covariantly modifies the conservation equations in the dark sector.
Using Cosmic Microwave Background (CMB), Baryon Acoustic 
Oscillations (BAO), and Supernovae type Ia (SnIa) data, we show that the Hubble tension can be alleviated due to the additional radiation, while the $\sigma_8$ tension present in the $\Lambda$-Cold-Dark-Matter model can be eased by the weaker galaxy clustering that occurs in these interacting models. Furthermore, we show that, while CMB+BAO+SnIa data put only upper bounds on the coupling strength, adding low-redshift data in the form of a constraint on the parameter $S_8$ strongly favours nonvanishing values of the interaction parameters. Our findings are in line with other results in the literature that could signal a universal trend of the momentum exchange among the dark sector.
\end{abstract}

\date{\today}
\maketitle
\section{Introduction}
\label{introsec}

The quality of the available cosmological data has improved dramatically in the last decades, making cosmology undergo a transition from an {\it order of magnitude} towards a precision science. Nowadays, we have data from a wide variety of sources such as Supernovae type Ia (SnIa)~\cite{Riess:1998cb,Perlmutter:1998np,WoodVasey:2007jb,Suzuki:2011hu,Scolnic:2017caz}, Baryon Acoustic Oscillations (BAO) from galaxy surveys~\cite{Eisenstein:2005su,2011MNRAS.416.3017B,Ross:2014qpa,Alam:2016hwk} and Lyman-$\alpha$ data~\cite{Blomqvist:2019rah,Agathe:2019vsu}, 
Cosmic Chronometers~\cite{Jimenez:2001gg,Stern:2009ep}, 
the anisotropies of the Cosmic Microwave Background (CMB)~\cite{Spergel:2003cb,Aghanim:2018eyx}, weak lensing measurements~\cite{Hildebrandt:2016iqg,Abbott:2017wau}, cluster counts, and structure formation or 
galaxy distribution~\cite{Tegmark:2006az,Blake:2011rj}. 

As these cosmological observations have increased both in quantity and quality, cosmologists have been faced with new 
challenges. On the one hand, the improved precision of the data made it necessary to have a better control on the precision of theoretical predictions from physical effects, foregrounds, etc. On the other hand, the variety of observables from different sources started showing some tensions among the data (see e.g., Refs.~\cite{Verde:2019ivm,Perivolaropoulos:2021jda}). 
In the standard $\Lambda$-Cold-Dark-Matter 
($\Lambda$CDM) model, for example, the value of today's Hubble expansion rate is constrained from the Planck CMB data as $H_{0}=(67.27\pm0.60)\Hunits$~\cite{Aghanim:2018eyx}. 
This is in tension with the bound
$H_0=(74.03\pm1.42)\Hunits$ obtained from supernovae calibrated on Cepheids (through the cosmic ladder) measured by the SH0ES collaboration~\cite{Riess:2018uxu} with a statistical significance of more than $3\sigma$. 
An independent local measurement by the H0LiCOW collaboration~\cite{Wong:2019kwg} using lensed quasars gives $H_{0}=73.3^{+1.7}_{-1.8}\Hunits$, that is also in tension with the Planck value and follows the trend of a statistically significant higher value inferred from local measurements. An exception to this trend is the Carnegie-Chicago collaboration~\cite{Freedman:2019jwv} that used a calibration of the Tip of the Red Giant Branch to measure $H_{0}=(69.8\pm2.5)\Hunits$ that is half way the Planck and SH0ES values and consistent with both. More extensive compilations of local measurements of the Hubble constant can be found 
in Refs.~\cite{Verde:2019ivm,DiValentino:2021izs,Perivolaropoulos:2021jda} (see also Fig.~\ref{Fig:dataH0} below). This discrepancy in the values of $H_0$ is arguably the most pressing tension in current cosmology, but it is not the only one. Another commonly quoted source of tension in cosmological data comes from the amplitude of matter perturbations, usually described by the parameter $\sigma_8$, as measured with CMB data and low-redshift probes like shear-lensing~\cite{Heymans:2012gg,Hildebrandt:2016iqg,Abbott:2017wau} and redshift-space distortions~\cite{Samushia:2013yga,Macaulay:2013swa}, albeit with a significance weaker than that of the $H_0$. 

It is presently unclear whether both the $H_0$ and  $\sigma_8$ tensions originate from unknown systematics or improper calibration/modelling or they truly indicate physics beyond the standard model. Regarding the $H_0$ tension, it has been recently claimed in Ref.~\cite{Mortsell:2021nzg} that a miscalibration of the cepheids colour might crucially impact the final value of $H_0$ so that it could even be compatible with the Planck value. Concerning the $\sigma_8$ tension, it has also been argued in Ref.~\cite{Blanchard:2021dwr} that it could be due to a difference in the bias parameter as inferred from observations and as derived from simulations. 

The problem has been scrutinised and exacerbated in recent publications~\cite{Verde:2019ivm,Riess:2019cxk,Freedman:2019jwv,Vagnozzi:2019ezj}. It is nevertheless worth pursuing alternatives to the $\Lambda$CDM model and studying whether there are models of dark energy (DE) or dark matter (DM) able to naturally alleviate the  tensions. In the presence of interactions between DE and DM, the cosmic expansion and growth histories are generally different  from those in the $\Lambda$CDM. Indeed, there are phenomenological models of coupled 
DE and DM that can ease the $H_0$ and (or) $\sigma_8$ tensions~\cite{DiValentino:2017iww,Yang:2018euj,Pan:2019gop,DiValentino:2019jae}. 

In most of the phenomenological approaches to the coupled DE and DM taken in the literature, the interacting terms between DE and DM are added to the background continuity equations by hand~\cite{Dalal:2001dt,Chimento:2003iea,Wang:2005jx,Wei:2006ut,Amendola:2006dg,Guo:2007zk,Valiviita:2008iv,Salvatelli:2014zta,Kumar:2016zpg,Salzano:2021zxk}. In this case, there is an ambiguity on how to promote the continuity equations to fully covariant forms~\cite{Tamanini:2015iia}. The problem manifests itself at the level of perturbations, in that the perturbation equations do not unambiguously follow from a single covariant action.

If we consider nonminimally coupled scalar-tensor theories like Brans-Dicke theories~\cite{Brans:1961sx}, the corresponding actions 
can be transformed to those in the Einstein frame with a standard Einstein-Hilbert term~\cite{Maeda:1988ab}. 
This transformation gives rise to a coupling between the DE scalar field and CDM~\cite{Wetterich:1994bg,Amendola:1999qq,Amendola:1999qq}. 
In such theories, the continuity equations of DE and DM in the Einstein frame are expressed as covariant forms, so they are free from the aforementioned problem. Since the coupling modifies the distance to the last scattering surface, this allows a possibility for reducing the $H_0$ tension~\cite{Ade:2015rim,Gomez-Valent:2020mqn,Ballesteros:2020sik,Braglia:2020auw}. However, this interaction corresponds to an energy transfer between DE and DM. In this case, it is difficult to address the problem of the $\sigma_8$ tension due to the enhancement of matter perturbations in comparison to the $\Lambda$CDM~\cite{Amendola:2003wa}. On the other hand, the presence of a momentum transfer can lead to the suppressed cosmic growth rate~\cite{Amendola:2020ldb}.

For a scalar field $\phi$ coupled to the CDM four-velocity $u_c^{\mu}$ weighed by a scalar quantity $Z=u_c^{\mu} \partial_{\mu}\phi$~\cite{Pourtsidou:2013nha,Boehmer:2015sha,Skordis:2015yra,Koivisto:2015qua,Pourtsidou:2016ico,Dutta:2017kch,Linton:2017ged,Kase:2019veo,Kase:2019mox,Chamings:2019kcl,Kase:2020hst}, the interacting Lagrangian like $f \propto Z^2$ suppresses the growth of matter perturbations through the momentum exchange between $\phi$ and CDM. There are also interacting models based on vector-tensor theories in which the vector field $A_{\mu}$ coupled to the CDM four-velocity of the form $u_c^{\mu}A_{\mu}$ leads to weak gravitational couplings for linear cosmological perturbations~\cite{DeFelice:2020icf}. In these models, the momentum exchange between DE and DM can alleviate the $\sigma_8$ tension~\cite{Pourtsidou:2016ico}.

The scalar or vector fields are not the only 
possibility for realizing late-time cosmic 
acceleration, but there are also fluid 
models of DE interacting with DM. 
In Ref.~\cite{Asghari:2019qld}, 
the authors proposed 
interacting fluid models of DE and DM in which 
their continuity equations are modified 
at the covariant 
level (see also Refs.~\cite{Jimenez:2020ysu,Figueruelo:2021elm}). 
The momentum 
exchange is quantified by the difference 
between the DE and CDM (or baryon) four 
velocities, with a coupling constant 
$\alpha$. In this model, the growth of 
CDM perturbations is suppressed by the 
presence of such a coupling~\cite{Figueruelo:2021elm}.
In Ref.~\cite{Jimenez:2020npm}, the present authors 
proposed perfect-fluid models of coupled DE and DM 
with a covariant action. The interaction 
is weighed by the scalar product 
$Z=g_{\mu \nu}u_c^{\mu}u_d^{\mu}$ 
together with a coupling constant $b$, 
where $u_c^{\mu}$ and $u_d^{\mu}$ are CDM 
and DE four velocities respectively.
In this setup, it is possible to construct 
an explicit interacting model in which DE initially works 
as a dark radiation and it behaves as a  cosmological constant at late times.
The momentum exchange between DE and CDM again leads to the suppressed growth of CDM perturbations. 
The characteristic property of these two classes of coupled fluid models is that only the perturbations are sensitive to the 
interaction, while the background evolution is oblivious to it. 

In this paper, we will place observational constraints on the two classes of interacting fluid models of DE and DM mentioned above. 
Besides the SnIa, BAO, and CMB data, we also implement the data of Sunyaev-Zeldovich (SZ) galaxy counts 
in the analysis. We will show that the 
$\sigma_8$ tension is alleviated 
by the momentum transfer between DE 
and DM, while the $H_0$ tension is not 
significantly eased with the SZ data taken 
into account.
In particular, inclusion of the SZ data favours nonvanishing values of the interaction parameters $\alpha$ and $b$. 
This suggests a signature for
the presence of momentum exchanges between 
the dark sectors.

\section{Dark elastic interactions}

We will start by briefly introducing the interacting models of DE and DM to be used in the subsequent analysis. These models have been more extensively discussed in the literature, so we will restrict to describe their main properties here and refer to the corresponding references~\cite{Asghari:2019qld,Figueruelo:2021elm,Jimenez:2020npm} 
for more details. 

A defining feature of these models is that the interaction is governed by the relative motion of the interacting components and only affects the Euler equations, at least in the realm of linear cosmological perturbation theory. Neither the background evolution nor the continuity equations of the linear density contrasts are modified. For this reason we refer to these interactions as {\it elastic}, with an admittedly  abuse of language, because there is no energy transfer and only momentum is exchanged. However, while the background is completely blind to the interaction, the density contrast feels it indirectly via the source term proportional to the velocity. The general physical mechanism is then that the interaction proportional to the relative motion of the fluids provides the DM component a drag that prevents the (linear) clustering to occur as efficiently as in the uncoupled situation. 
This effect causes a suppressed matter power spectrum that could eventually alleviate the $\sigma_8$ tension. Since the interaction is proportional to the relative motion and the fluids are assumed to be comoving on sufficiently large scales, the suppression will only take place for small scales. Since we will be concerned with the linear scales only, these small scales refer to sub-Hubble modes but above the non-linear scale. 

We will focus on scalar perturbations on top of a spatially flat Friedmann-Lema\^{i}tre-Robertson-Walker (FLRW) background, 
so the metric is given by the line element
\be
{\rm d}s^2=-(1+2\xi) {\rm d}t^2
+2 \partial_i \chi {\rm d}t {\rm d}x^i
+a^2(t) \Big[ (1+2\zeta) \delta_{ij}
+2\partial_i \partial_j E \Big] {\rm d}x^i {\rm d}x^j\,,
\label{permet}
\ee
with $a$ the scale factor and $\xi$, $\chi$, $\zeta$ and $E$ the scalar perturbations. Concerning the matter sector, we assume the standard components conformed by photons, neutrinos and baryons together with a dark sector containing pressureless CDM with energy-momentum tensor 
\be
T_c^{\mu\nu}=\rho_c u^\mu_c u^\nu_c\,,
\ee
where $\rho_c$ is the CDM density and a DE component whose energy-momentum tensor is
\be
T_d^{\mu\nu}=(\rho_d+P_d) u^\mu_d u^\nu_d+P_d g^{\mu\nu}\,,
\ee
where $\rho_d$ and $P_d$ are the density and pressure of DE, 
respectively. We decompose the fluid variables $\rho_d$ and $P_d$ 
into background and perturbed 
values as $\rho_d=\bar{\rho}_d+\delta \rho_d$ and 
$P_d=\bar{P}_d+\delta P_d$. 
The background DE dynamics can be characterised by the equation of state parameter $w_d=\bar{P}_d/\bar{\rho}_d$. Furthermore, we introduce the sound speed $c_d^2 =\delta P_d/\delta \rho_d$ associated with the propagation of perturbations.

We will mostly employ the Newtonian gauge defined by $\chi=E=0$ and the Newtonian potentials $\Psi=\xi$ and $\Phi=-\zeta$ for the physical interpretation of the results. 
Since we will not consider anisotropic stress, we have that $\Phi=\Psi$ for the relevant scales. However, the synchronous gauge will also be used for the implementation of the interacting models into CAMB. After introducing the general setup, we proceed to introducing the two models that we will analyse in this work.

\subsection{Velocity-entrainment coupling}
\label{models}

The first model that we will consider was introduced in Ref.~\cite{Jimenez:2020npm} and consists of two perfect fluids describing DE and DM with a coupling that gives a rise to new terms in the equations of motion depending on the relative motion. 
To construct the interaction, we first define the quantity
\be
Z_{ij}\equiv g_{\mu\nu}u^\mu_i u^\nu_j
\ee
for two fluids $i$ and $j$. This quantity closely resembles the so-called entrainment of two fluids that can be written as $Z_{ij}^{\rm ent}= n_i n_j Z_{ij}$, with $n_i$ the respective number density of the two fluids.\footnote{The entrainment $Z_{ij}$ was used in Ref.~\cite{Ballesteros:2013nwa} to build the effective field theory of several interacting fluids. In their general scenario, they introduced another higher-order object of relevance when having more than three interacting fluids. Our scenario is thus a particular case of the system studied in  Ref.~\cite{Ballesteros:2013nwa} and we could alternatively use their formalism instead of our starting action \eqref{action}, although the resulting equations for the cosmological perturbations are not expected to depend on the employed formalism.} Crucially, $Z$ is normalised so that $Z_{ii}=-1$ and it only depends on the velocities, unlike the entrainment that also depends on the abundances. For this reason we refer to $Z_{ij}$ as the {\it velocity-entrainment}. The fact that the interaction depends on this precise combination and not the entrainment is crucial for the interaction not to affect the background nor the continuity equations as we will see next. 
Since our two-fluid system is conformed by CDM and DE, we will use $Z\equiv Z_{cd}$ to alleviate the notation. The system can be described by the following action \cite{Jimenez:2020npm}
\be
{\cal S} =
\frac{M_{\rm pl}^2}{2}\int {\rm d}^4 x \sqrt{-g}\,R
-\sum_{I=c,d,b,r}\int {\rm d}^{4}x 
\Big[\sqrt{-g}\,\rho_I(n_I)+ J_I^{\mu} \partial_{\mu} \ell_I \Big]
+\int {\rm d}^4x\sqrt{-g}\,f(Z)\,,
\label{action}
\ee
where the first term is the 
standard Einstein-Hilbert action with the reduced 
Planck mass $M_{\rm pl}$ and the determinant $g$ 
of metric tensor $g_{\mu \nu}$, and the second one describes the cosmological perfect fluids\footnote{Of course, deviations from the perfect-fluid approximation arise on sufficiently small scales for instance in the form of viscosity and heat conduction in the photons-baryons fluid before recombination that give rise to the Silk damping. The effects of these imperfections are included in the full numerical analysis. However, for the derivation of the coupled DE and DM equations, these are not relevant so we omit them for clarity.} with the Schutz-Sorkin action \cite{Schutz:1977df,Brown:1992kc,DeFelice:2009bx}. The different fluids correspond to different choices of the function $\rho_I(n_I)$.
The subscripts $c,d,b,r$ stand for CDM, DE, baryons, and radiation, respectively. The fields $\ell_I$ act as Lagrange multipliers ensuring the conservation of  currents $J^\mu_I$ that are related to the 
corresponding number density as
\be
n_I=\sqrt{\frac{g_{\mu\nu} J^\mu_I J^\nu_I}{g}}.
\ee
The interaction is encoded in the last piece 
of Eq.~\eqref{action} given in terms of the arbitrary function $f(Z)$ that can always be chosen to satisfy
\be
f(-1)=0\,,
\ee
at the expense of shifting the value of the cosmological constant. 
This condition guarantees that the interaction does not appear in the background equations of motion that read as usual
\ba
& &
3M_{\rm pl}^2 H^2 
= \sum_{I=c,d,b,r} \rho_I\,,\\
& &
M_{\rm pl}^2 \left( 2\dot{H}+3H^2 
\right) = -\sum_{I=c,d,b,r} P_I\,,\\
& &
\dot{\rho}_I+3H \left( \rho_I+P_I 
\right)=0\,,
\ea
where $P_I=n_I \rho_{I,n_I}-\rho_I$ is the fluid 
pressure with the notation $\rho_{I,n_I}=
\partial \rho_I/\partial n_I$ and
$H=\dot{a}/a$ is the Hubble expansion rate, where a dot represents the derivative with respect to $t$. Here and in the following, we omit a bar from the background quantities. 
The equation of state parameter for each matter component is defined by $w_I=P_I/\rho_I$.

The derivation of the full linear perturbation equations of motion has been worked out and analysed 
in detail in Ref.~\cite{Jimenez:2020npm} where we refer for more details and here we will only give the most relevant equations. The linear perturbation equations in the dark sector are only sensitive to the interaction through $f_{,Z}$ evaluated on the background. Since we have that $Z=-1$ at the background level, the interaction only introduces an additional constant parameter that we denote
\be
b\equiv (f_{,Z})_{|_{Z=-1}}\,.
\ee
This is a remarkable property since this whole class of interacting models can be universally described, at linear order in cosmological perturbations, by a single parameter regardless the functional form of $f(Z)$. From a model-building perspective, this is a very valuable feature because additional parameterisations or Ans\"atze to describe the interaction are not necessary and, consequently, it reinforces the predictive power of this class of models. 

The DM component is taken as the standard pressureless fluid of the $\Lambda$CDM model, so it will be described by $\rho_c\propto n_c$. Thus, we only need to specify the DE sector. 
Following the motivation given 
in Ref.~\cite{Jimenez:2020npm}, we assume that the DE sector is described by
\be
\label{rhod0}
\rho_d=\rho_\Lambda \left(1+r_d n_d^{1+c_d^2}\right)=\rho_\Lambda
\left[1+\frac{\Omega_{df}}{\Omega_\Lambda} 
a^{-3(1+c_d^2)}\right] 
\equiv \rho_{\Lambda}+
\rho_{df} \,,
\ee
where $\rho_\Lambda$, $r_d$, $c_d^2$, $\Omega_{\Lambda}$, and $\Omega_{df}$ are positive constant parameters, out of which only three of them are independent. In the last equality we have explicitly split the density $\rho_d$ into a cosmological constant $\rho_\Lambda$ 
and a dark fluid described by $\rho_{df}=\rho_{\Lambda} r_d n_d^{1+c_d^2}$, so it has equation of state $w_{df}=c_d^2$, 
with today's density parameters $\Omega_{\Lambda}$ 
and $\Omega_{df}$ respectively.
The corresponding DE pressure and equation of state parameter 
are given, respectively, by 
\be
P_d=-\rho_{\Lambda}+c_d^2 
\rho_{df}\,,\qquad
w_d=\frac{-\rho_{\Lambda}
+c_d^2 \rho_{df}}{\rho_{\Lambda}+\rho_{df}}\,.
\label{eq:EoSd}
\ee
At early times where 
$\rho_{df} \gg \rho_{\Lambda}$, 
the equation of state $w_d$ reduces to $c_d^2$.
In particular, for $c_d^2=1/3$, it works as a dark radiation. After $\rho_{\Lambda}$ dominates over $\rho_{df}$ at late times, $w_d$ approaches $-1$. This is of course consistent with our splitting of the DE fluid into a cosmological constant and a perfect fluid with equation of state $w_{df}=c_d^2$.

In the Newtonian gauge, the density contrasts $\delta_{I}$ and velocity potentials $\theta_{I}$ of CDM 
and DE (with the subscripts 
$I=c, d$, respectively) obey the following 
differential equations \cite{Jimenez:2020npm}
\ba
& &
\delta_{c}'=
-\theta_{c}+3\Phi'\,,
\label{delceq}\\
& &
\delta_{d}'=-3{\cal H} 
\left( c_d^2-w_d \right) 
\delta_{d}+3(1+w_d)\Phi'
-(1+w_d)\theta_{d}\,,
\label{deldeq}\\
& &
\theta_{c}'=-{\cal H}
\theta_{c}+k^2 \Phi 
+b\frac{3 {\cal H} (1+w_d) \rho_d 
[\theta_{c}-(1+c_d^2) \theta_{d}]
-k^2 c_d^2 \rho_d \delta_{d}}
{(1+w_d)\rho_d(\rho_c-b)-b\rho_c}\,,
\label{thetaceq} \\
& &
\theta_{d}'={\cal H} (3c_d^2-1) \theta_{d}+k^2 \Phi 
+\frac{\rho_c[k^2 c_d^2 \rho_d \delta_{d}+
3 {\cal H} b\{(1+c_d^2) \theta_{d}-\theta_{c}\}]
-k^2 b c_d^2 \rho_d \delta_{d}}
 {(1+w_d)\rho_d(\rho_c-b)-b\rho_c}\,,
 \label{thetadeq}
\ea
where a prime represents the derivative with respect to the conformal time $\tau=\int a^{-1}{\rm d}t$, and ${\cal H}=aH$. Since the DE sector that we are assuming is simply a cosmological constant plus a perfect fluid, the interaction with CDM is actually oblivious to the cosmological constant sector (other than its effects through the background evolution). Thus, we can alternatively rewrite the above system of equations governing the dark sector as standard $\Lambda$CDM with an additional dark perfect fluid with a constant equation of state parameter given by $c_d^2$. At the background level, this latter dark fluid obeys the continuity equation
\be
\rho'_{df}+3{\cal H}(1+w) 
\rho_{df}=0\,,
\ee
where $w=c_d^2$. The relation 
between $w$ and $w_d$ is 
given by 
\be
(1+w) \rho_{df}=(1+w_d) \rho_d\,.
\ee
The density contrast of the dark fluid is defined as usual $\delta_{df}=\delta\rho_{df}/\rho_{df}$. 
Since the cosmological constant piece of the DE fluid does not contribute to the density perturbation, we can write
\be
\delta_{df}=\frac{\delta \rho_d}{\rho_{df}}=
\frac{\rho_d}{\rho_{df}}\delta_d=
\frac{\rho_{\Lambda}+\rho_{df}}{\rho_{df}}\delta_d\,,
\ee
which relates the density contrasts of the dark fluid and that of the total DE component. 
Obviously, the difference can be entirely attributed to a different normalisation of the truly fluctuating piece. Furthermore, the velocity potentials are simply related by $\theta_{df}=\theta_d$. 
Then, the perturbation 
Eqs.~(\ref{delceq})-(\ref{thetadeq}) can be equivalently expressed as the following system
\ba
& &
\delta_{c}'=-\theta_{c}+3\Phi'\,,
\label{delceq2}\\
& &
\delta_{df}'=
3(1+w)\Phi'-(1+w)\theta_{df}\,,
\label{deldeq2}\\
& &
\theta_{c}'=-{\cal H}\theta_{c}
+k^2 \Phi 
+b\frac{3 {\cal H} (1+w) \rho_{df} 
[\theta_{c}-(1+c_d^2) \theta_{df}]
-k^2  c_d^2 \rho_{df} \delta_{df}}
{(1+w)\rho_{df}(\rho_c-b)-b\rho_c}\,,
\label{thetaceq2} \\
& &
\theta_{df}'={\cal H} (3c_d^2-1) \theta_{df}+k^2 \Phi 
+\frac{\rho_c[k^2 c_d^2 \rho_{df} \delta_{df}+
3 {\cal H} b\{(1+c_d^2) \theta_{df}-\theta_{c}\}]
-k^2 b c_d^2 \rho_{df} \delta_{df}}
 {(1+w)\rho_{df}(\rho_c-b)-b\rho_c}\,.
 \label{thetadeq2}
 \ea
Although not obvious from the above form of the equations, 
the effect of the interaction depends on the relative motion 
of two fluids so that it vanishes if the two fluids have the same velocity (see Ref.~\cite{Jimenez:2020npm}). What it is apparent from the equations is that only the Euler equations depend on the interaction, whose effect is driven by the parameter $b$. For later convenience, we will introduce a  dimensionless version of the interaction parameter by normalising to today's critical density 
$\rho_{\rm crit}=3M_{\rm pl}^2 H_0^2$ as
\be
b\rightarrow \frac{b}{\rho_{\rm crit}}\,.
\ee

We have implemented both sets of equations (\ref{delceq})-(\ref{thetadeq}) and 
(\ref{delceq2})-(\ref{thetadeq2}) 
into the CAMB code\footnote{https://camb.info} and corroborated that the results are in excellent agreement with each other. However, we find the second approach  cleaner and more physically transparent when interpreting the results below. 
As in Ref.~\cite{Jimenez:2020npm}, we will further assume that the additional perfect fluid in the dark sector corresponds to a dark radiation component with $c_d^2=1/3$. Expressing $\Omega_{df}$ in Eq.~(\ref{rhod0}) 
as $\Omega_{\rm DR}$, we have
\be
\label{rhod1}
\rho_d=\rho_\Lambda\left(1+\frac{\Omega_{\rm DR}}{\Omega_\Lambda} a^{-4}\right).
\ee
This assumption is motivated to have a scaling evolution in the early Universe. In this case the background evolution is not sensitive to the redshift at which we set the initial conditions, but it only depends on the fractional abundance of dark radiation with respect to ordinary radiation. 
As explained in detail 
in Ref.~\cite{Jimenez:2020npm}, the interaction reduces the growth of structures on small scales so that the overall suppression is controlled by $b$, while $\Omegadr$ determines the scales that undergo the suppression. We will confirm this result below from the full numerical solution.

To finish the brief introduction of this model, we will discuss an interesting observation regarding the employed DE fluid that is composed of dark radiation and a cosmological constant. 
This is in fact what is obtained for a null fluid 
whose energy momentum tensor is given by
\be
T_{\mu\nu}=\left( \rho+P \right)\ell_\mu \ell_\nu
+P g_{\mu\nu}\,,
\ee
with $\ell_\mu\ell^\mu=0$ in the FLRW universe. 
In fact, the conservation equations for this fluid lead to $P=P_{0}$ and $\rho=\rho_0a^{-4}-P_0$, where $P_0$ is a constant.
This energy density exactly corresponds to a combination of the cosmological constant and radiation as we are assuming for our DE sector. The similarities however end there, although it would be interesting to explore to what extent we can push this analogy as well as the inclusion of null fluids in our DE sector.

\subsection{Covariantised dark Thomson-like scattering}
\label{models_elastic}

The second class of models that we will consider is based on an interaction introduced at the level of conservation 
equations in the following manner~\cite{Asghari:2019qld}:
\begin{eqnarray}
\nabla_\mu T^{\mu\nu}_c &=&\alpha(u^\nu_c-u^\nu_d)\,,\\
\nabla_\mu T^{\mu\nu}_d&=&
-\alpha(u^\nu_c-u^\nu_d)\,,
\label{eq:Defintalpha}
\end{eqnarray}
where $\alpha$ describes the strength of the interaction.  
In principle, it could be some function of the fluid variables, but we follow here the same assumptions as 
in Ref.~\cite{Asghari:2019qld} and consider $\alpha$ to be constant. We see here the need for additional assumptions as compared to the previous model where the interaction was described by a single constant parameter in a natural way. Also in this case it will be convenient to work with a dimensionless coupling constant normalised as
\be
\alpha \rightarrow 
\rho_{\rm crit} H_0 
\alpha\,.
\ee
As in the previous case, the most important property of the interaction is the absence of energy-exchange\footnote{Strictly speaking, this is true up to first order in cosmological perturbations and provided the two fluids share the same large-scale rest frame.}, 
so only the perturbed Euler equations receive a correction proportional to $\alpha$. This is already apparent from the form of the interaction in Eq.~\eqref{eq:Defintalpha} that depends on the relative velocity of two components. 
Since the two components are comoving on large scales, the background equations are not modified, while the linear perturbation
equations of motion in the Newtonian 
gauge are given by \cite{Asghari:2019qld,Figueruelo:2021elm}
\begin{eqnarray}
\label{eq:deltaDM}
\delta_{c}' &=& 
-\theta_{c}+3\Phi'\,,\\
\delta_{d}'&=&-3 \mathcal{H}( c_d^2-w_d) \delta_{d} +3(1+w_d)\Phi'  -(1+w_d)\left(1+9 \mathcal{H}^2
\frac{c_d^2-w_d}{k^2}\right)
\theta_d\;, \\
\label{eq:thetaDM}
\theta_c'&=&-\mathcal{H} \theta_c + k^2 \Phi + \Gamma(\theta_d-\theta_c)\;, \\
\label{eq:thetaDE}
\theta_d' &=&(3c_d^2-1) \mathcal{H}\theta_d+k^2\Phi +\frac{k^2 c_d^2}{1+w_d}\delta_d-\Gamma R_{cd}(\theta_d-\theta_c)\,,
\end{eqnarray}
where $c_d^2$ is the adiabatic sound speed squared, 
that we will assume to be $c_d^2=1$, and 
\begin{eqnarray}
\Gamma&\equiv& \alpha \frac{a}{\rho_c} \;, \\ \label{eq:Scoupling}
R_{cd} &\equiv&
\frac{\rho_c}{(1+w_d)\rho_d}\,. \label{eq:Rcoupling}
\end{eqnarray}
The quantity $\Gamma$ can be
interpreted as the interaction rate between CDM and DE, 
whereas $R_{cd}$ is the ratio between the CDM and DE background densities. We can see that the resulting new terms in 
Euler equations of the dark sector have the same form as those arising from a Thomson scattering (as e.g., in the baryon-photon fluid before recombination). 
The considered interaction 
in Eq.~\eqref{eq:Defintalpha} can then be interpreted as a covariantisation of such a scattering and thus our denomination for this interaction. The possible presence of a Thomson scattering involving the dark sector was discussed in Ref.~\cite{Simpson:2010vh}.

The resemblance of Eqs.~(\ref{eq:deltaDM})-(\ref{eq:thetaDE}) with those of the interacting model discussed in Sec.~\ref{models} is apparent. In both cases, the interaction only affects the Euler equations and it disappears on large scales due to the convergence of the DE and CDM rest frames. However, there are some differences between the two interacting scenarios. In the present model, for instance, CDM interacts with a DE fluid with an equation of state that we assume close to $-1$ but different from it. In the model of Sec.~\ref{models}, CDM is coupled to the dark radiation component of the DE fluid, while the accelerated expansion is driven by an exact cosmological constant. This latter model can be regarded as $\Lambda$CDM with a dark radiation sector that is coupled to CDM. It is nevertheless different from other coupled models of CDM and dark radiation because here dark radiation is regarded as an exact perfect fluid and we neglect any potential anisotropic stresses of higher moments in the would-be Boltzmann hierarchy of equations.

Since the model in this section does not modify the background, it is irrelevant for addressing 
the problem of the $H_0$ tension.
On the other hand, the model of Sec.~\ref{models} based on a velocity-entrainment coupling contains a dark radiation component that contributes to the background density, thus with a potential to alleviate the $H_0$ tension. Hence, in order to have a better understanding of the similarities and differences between the two interactions, we will allow  for a varying effective relativistic degrees of freedom parameter $N_{\rm eff}$ in our analysis of this second model, that amounts to including an additional radiation component.\footnote{Let us notice however that $N_{\rm eff}$ does not exactly mimic the effect of the perfect fluid dark radiation in the model with a velocity-entrainment coupling because $N_{\rm eff}$ assumes a proper radiation component.} The previous studies of this model in the literature \cite{Asghari:2019qld,Figueruelo:2021elm} have been performed with the standard value $N_{\rm eff}=3.046$ and kept fixed. Letting $N_{\rm eff}$ vary is in turn one of the motivations to include this model in the present work, so we will extend the existing observational limits to include the effect of a varying $N_{\rm eff}$. This extended version of the model exhibits apparent resemblance with the velocity-entrainment coupled model of Sec.~\ref{models}, so it is interesting to compare the results for both models to unveil some universal features of this class of interactions.

Thus, in the remaining of this paper, we will consider the following two interacting scenarios:
\begin{itemize}
    \item {\bf $b$CDM}$\equiv b+\Omega_{\rm DR}+\Lambda$CDM. This model contains the standard CDM fluid that interacts via a velocity-entrainment with a dark radiation fluid and DE is described by a pure cosmological constant. The parameters of this model are those of $\Lambda$CDM plus the interaction parameter $b$ (that universally captures the entire class of interactions) and the abundance of dark radiation $\Omega_{\rm DR}$, that is two more than $\Lambda$CDM.
    
    \item {\bf $\alpha$CDM}$\equiv \alpha+N_{\rm eff}+w$CDM. This model also contains the standard CDM fluid, but DE is dynamical with a constant equation of state $w$ and sound speed squared $\cs^2=1$. 
    The interaction is described by the additional parameter $\alpha$ and, in order to have a closer scenario to the previous model, we will also vary $N_{\rm eff}$. Thus, this model has three more parameters than those in $\Lambda$CDM.
\end{itemize}

Besides these two interacting models, we will also consider the standard $\Lambda$CDM for comparison. In some cases we will refer to an extended version of $\Lambda$CDM where we let $N_{\rm eff}$ be a free parameter and we will dub it $\Lambda$CDM$+N_{\rm eff}$. This will allow us to distinguish the importance of the interactions and extra radiation in the interacting models as well as comparing with previous literature. Furthermore, we should notice that the $\alpha$CDM model reduces to the standard $w$CDM plus varying $N_{\rm eff}$ for $\alpha=0$.

\section{Numerical results: Effects on the CMB and matter power spectra}
\label{sec:spectra}

Prior to obtaining observational constraints on the interacting models under consideration, we will first show most important effects on the evolution of the perturbations and, in particular, on the CMB and matter power spectra. We have used the implementation of covariantised Thompson-like interactions of Sec.~\ref{models_elastic} into modified versions of the publicly available codes CLASS~\cite{CLASS2011,CLASS2011JCAP} and CAMB~\cite{Lewis:1999bs,Howlett:2012mh} developed in Ref.~\cite{Figueruelo:2021elm}.

Concerning the $b$CDM model with a velocity-entrainment coupling, we have modified the CAMB code to include the effect of the interaction 
and extra dark radiation. 
As mentioned above, we have implemented 
modifications to CAMB following two equivalent approaches, namely: 
\begin{itemize}
    \item assuming the standard $\Lambda$CDM model and adding an extra radiation component that interacts with CDM via Eqs.~\eqref{delceq2}-\eqref{thetadeq2};
    \item assuming the DE sector with the equation of state given by Eq.~\eqref{eq:EoSd} and the interaction with CDM governed by Eqs.~\eqref{delceq}-\eqref{thetadeq}.
\end{itemize}
Both implementations are in perfect agreement between them and show similar performance. For this model we have not developed a modified version of CLASS. 

The initial conditions in $\alpha$CDM are set to be the same as in $\Lambda$CDM because the interaction is only relevant at low redshifts. 
On the other hand, the natural choice for $b$CDM is setting standard adiabatic initial conditions including the dark radiation component. 
We have checked that the evolution does not crucially depend on the chosen initial conditions (as long as they remain within reasonable limits) for the DE sector. For instance, setting vanishing initial perturbations for the DE component gives identical 
(late time) results as those derived with the more careful choice of adiabatic initial conditions. 
The reason is that the standard adiabatic mode is still the attractor for the evolution and the solutions quickly evolve towards it. In practice, we have used this property to set vanishing DE perturbations for the initial conditions. A potential subtle point with this class of models is the impossibility of using the residual gauge freedom of the synchronous gauge to set $\theta_c=0$ at all times. This, however, does not represent a problem in our case because we do not observe any dominant spurious gauge mode in our numerical solutions (see, 
e.g., Refs.~\cite{Jimenez:2020ysu,Figueruelo:2021elm}
for further details on this issue). 

\begin{figure}[!t]
\includegraphics[width=0.4\textwidth]{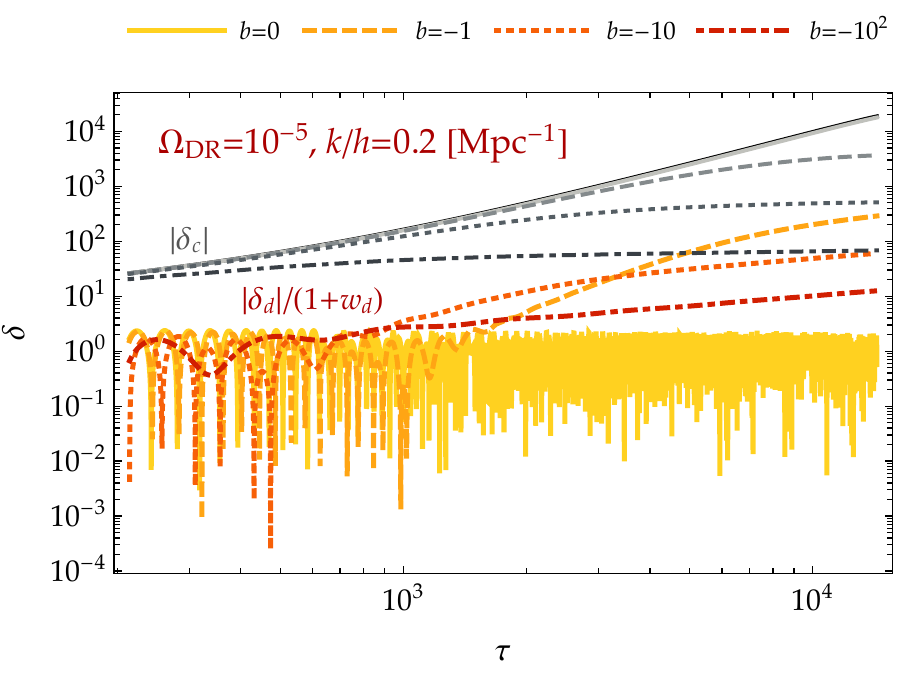}
\hspace{0.4cm}
\includegraphics[width=0.4\textwidth]{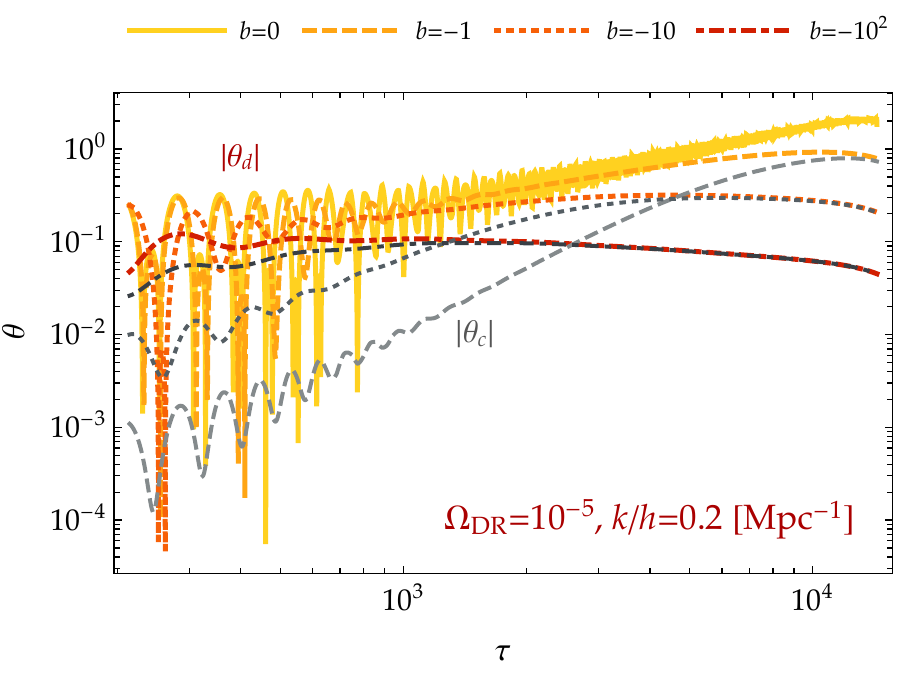} \\
\includegraphics[width=0.4\textwidth]{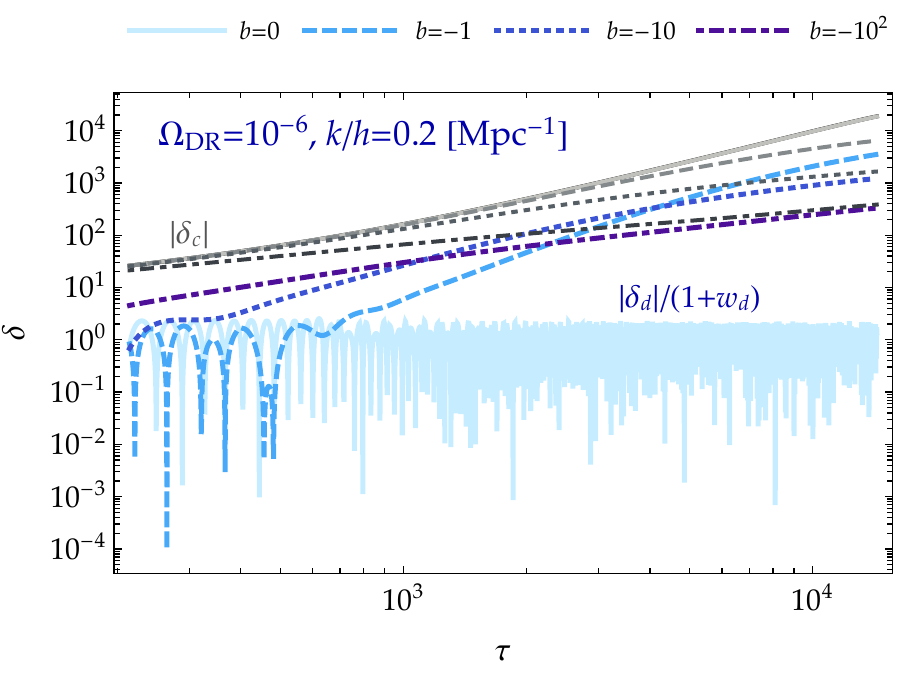}
\hspace{0.4cm}
\includegraphics[width=0.4\textwidth]{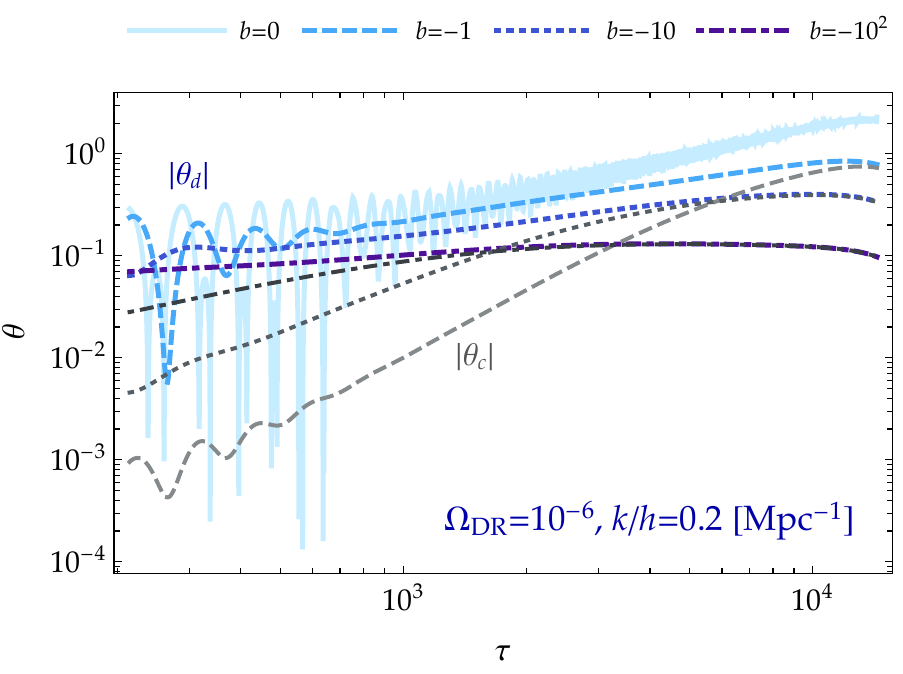} \\
\includegraphics[width=0.4\textwidth]{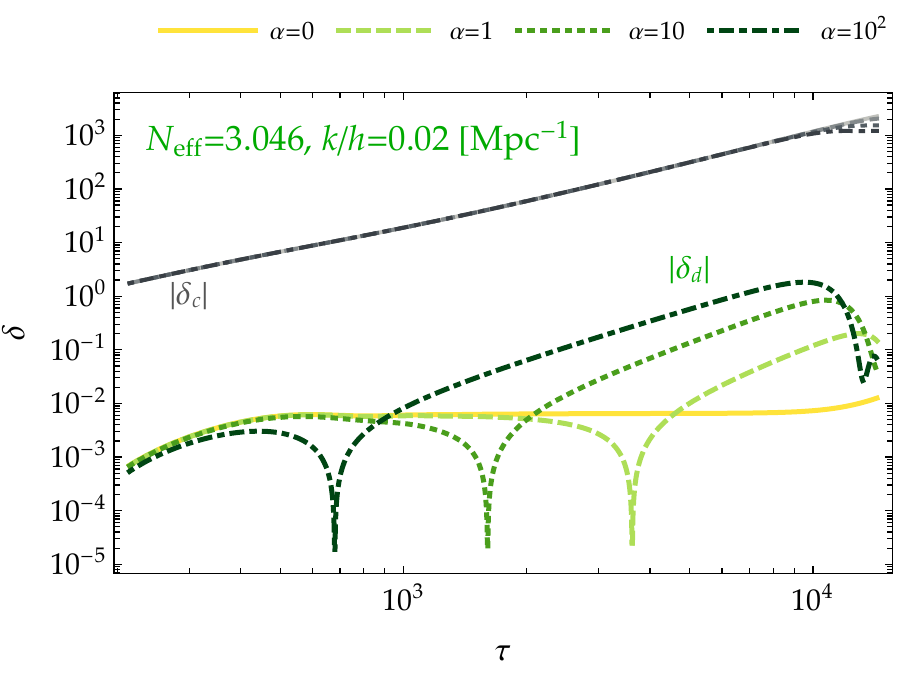}
\hspace{0.4cm}
\includegraphics[width=0.4\textwidth]{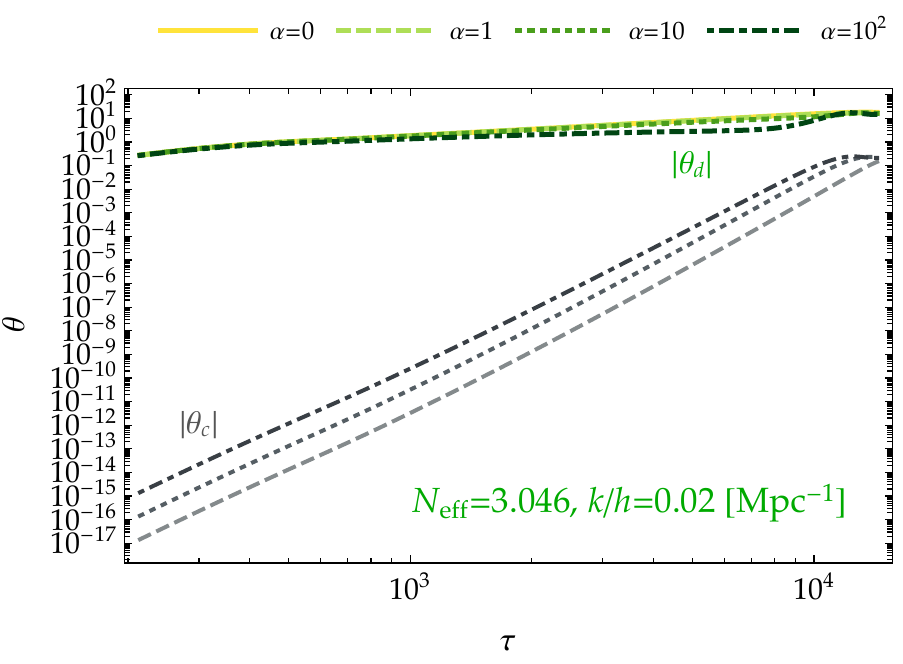} \\
\includegraphics[width=0.4\textwidth]{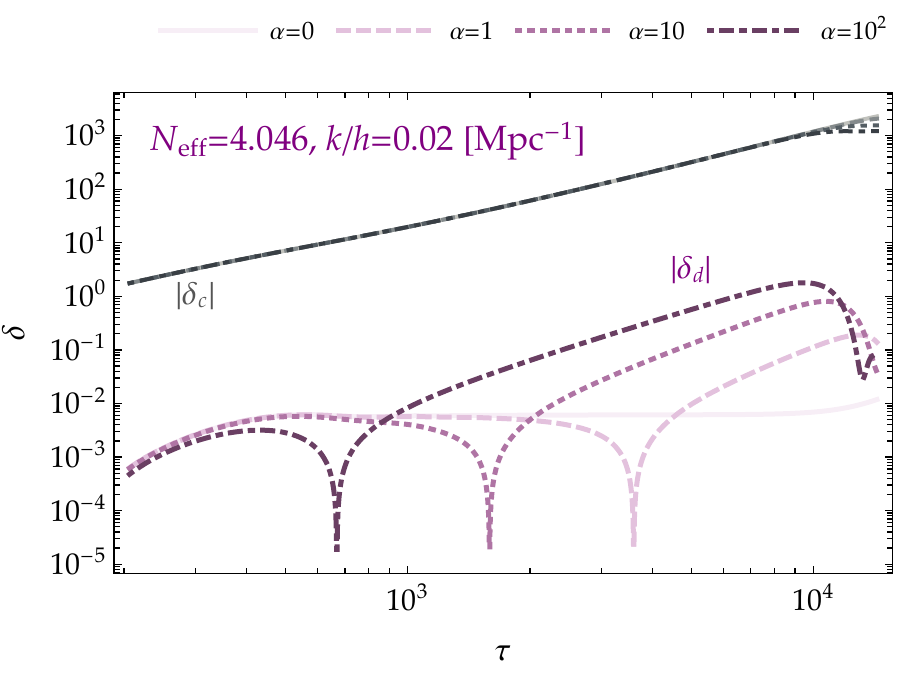}
\hspace{0.4cm}
\includegraphics[width=0.4\textwidth]{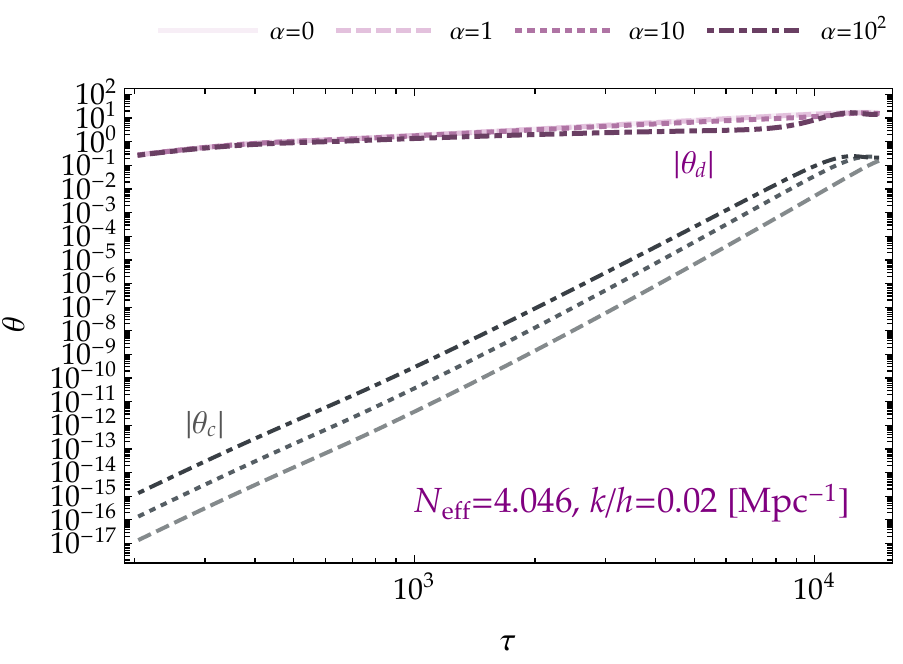} 
\caption{Evolution of the CDM and DE density contrasts and 
velocity potentials in the $b$CDM and $\alpha$CDM models for
different values of the parameters and wavenumbers. We can see how 
the growth of the CDM density contrast slows down at late times in both models, thus yielding a suppression in the growth of structures with respect to $\Lambda$CDM.
}
\label{fig:perturbations}
\end{figure}

In Fig.~\ref{fig:perturbations}, we plot the evolution of the perturbations in the dark sector for both models and some representative values of the parameters and scales. The main property to highlight is that, in both models, the CDM density contrast $\delta_c$ is subject to slower growth at late times that is driven by elastic interactions. The analytic estimations for the evolution of $\delta_c$ and $\delta_d$ were already performed in Refs.~\cite{Asghari:2019qld,Jimenez:2020npm} and they show good agreement with the obtained numerical results.

\begin{figure}[!t]
\includegraphics[width=0.95\textwidth,keepaspectratio]{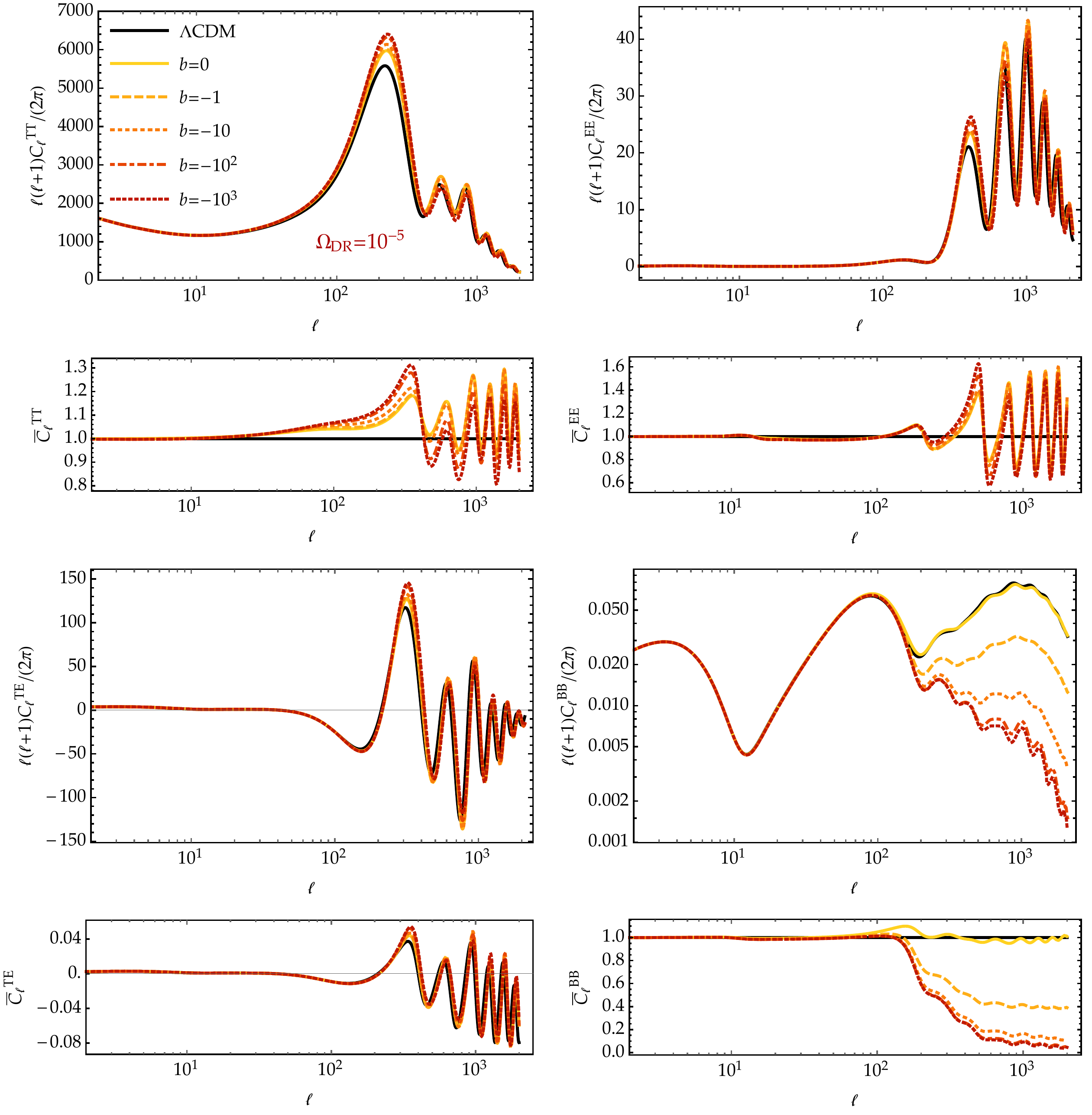}
\caption{Temperature and polarisation CMB angular power spectra 
for the $\Lambda$CDM and $b$CDM models with
$\Omega_{\rm DR}=10^{-5}$ and several values of 
the interaction parameter $b$. 
We also show the normalised values defined as $\bar{C}^{xx}_\ell=
C^{xx}_{\ell\ b{\rm CDM}}/C^{xx}_{\ell\ \Lambda {\rm CDM}}$, except 
$\bar{C}^{\rm TE}_\ell$ normalised with 
the temperature angular power spectrum.}
\label{fig:clsDR}
\end{figure}

\begin{figure}[!t]
\includegraphics[width=0.95\textwidth]{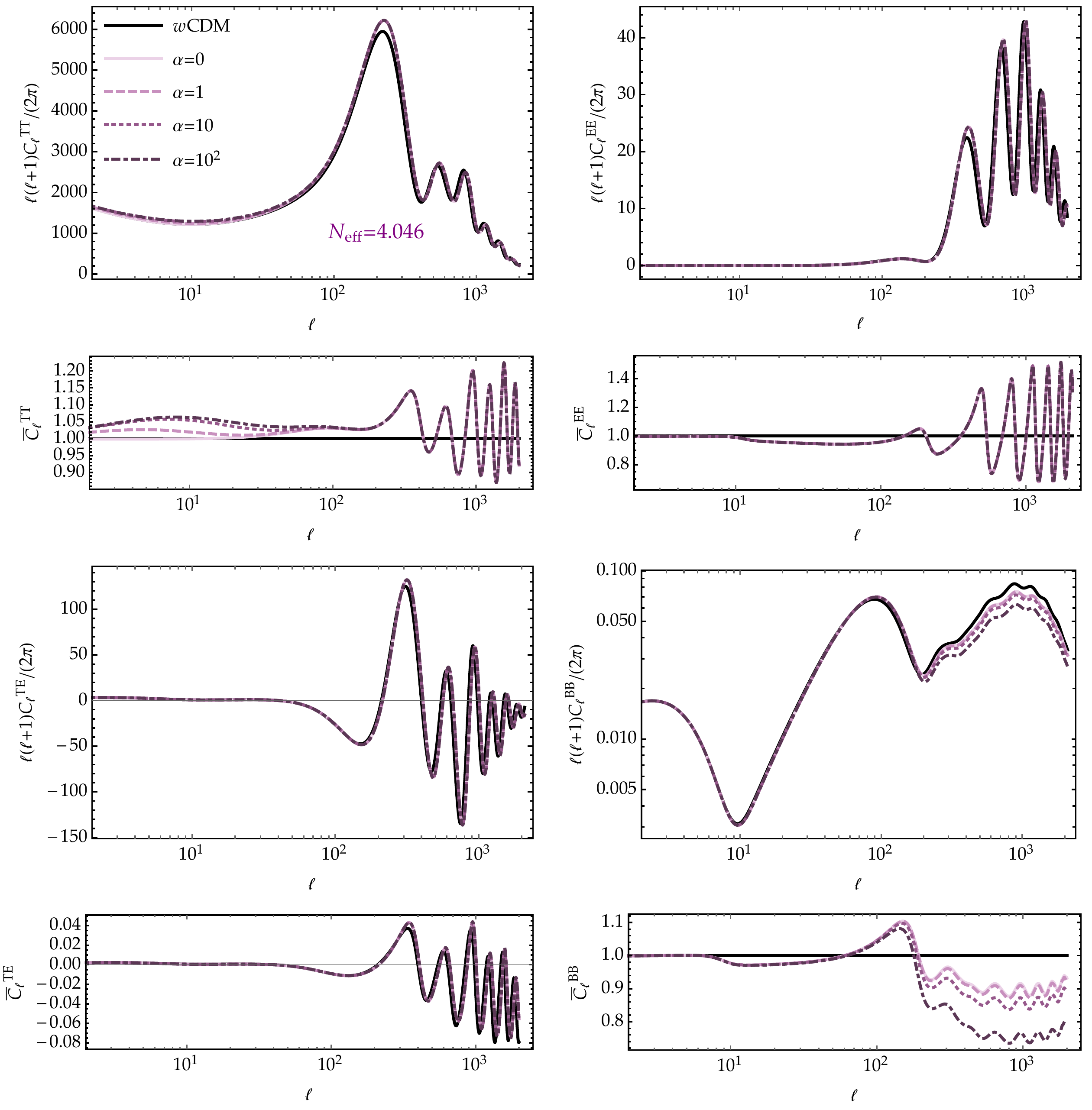}
\caption{Temperature and polarisation CMB angular power spectra in $\alpha$CDM for different values of the interacting parameter 
$\alpha$ and with $N_{\rm eff}=4.046$. 
The fiducial $w$CDM model with $w=-0.98$ is recovered 
for $\alpha=0$ and $N_{\rm eff}=3.046$. 
Normalised differences between $\alpha$CDM and $w$CDM
are included in the form $\bar{C}^{xx}_\ell=C^{xx}_{\ell\ \alpha{\rm CDM}}/C^{xx}_{\ell\ w{\rm CDM}}$, 
except 
$\bar{C}^{\rm TE}_\ell$ normalised 
with the temperature angular power spectrum.
}
\label{fig:clsEM}
\end{figure}

\begin{figure}[!t]
\includegraphics[width=0.48\textwidth]{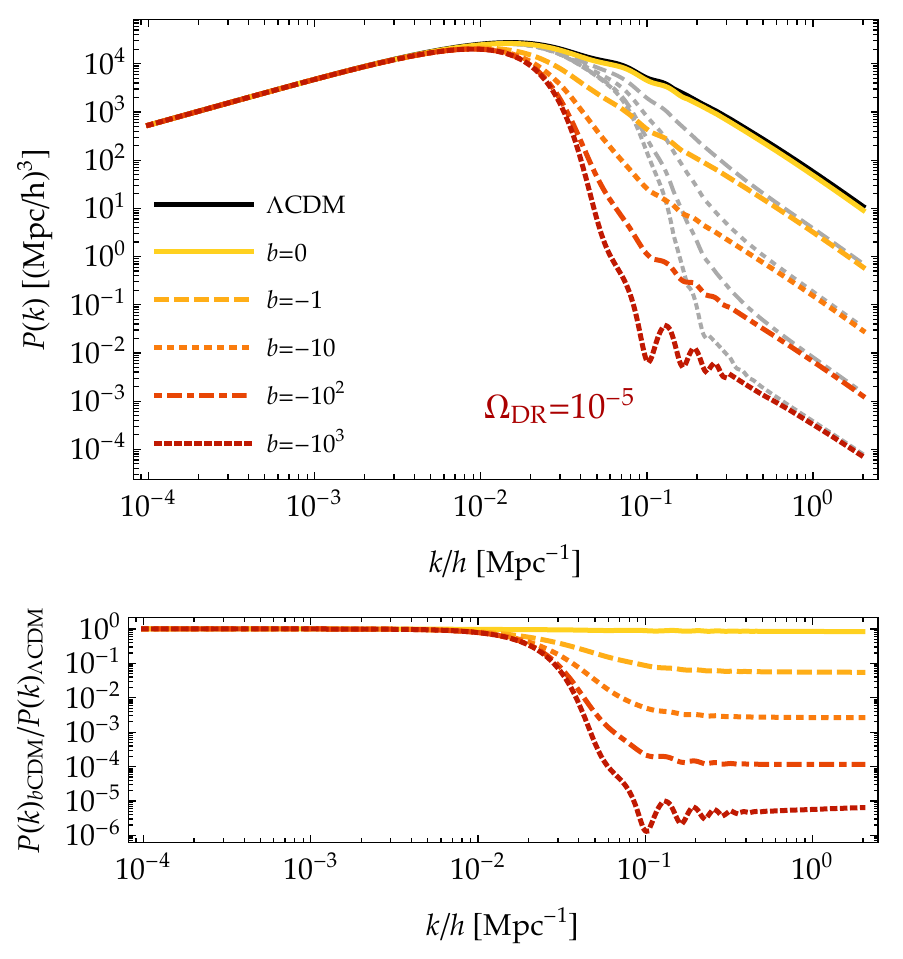}
\hspace{0.2cm}
\includegraphics[width=0.48\textwidth]{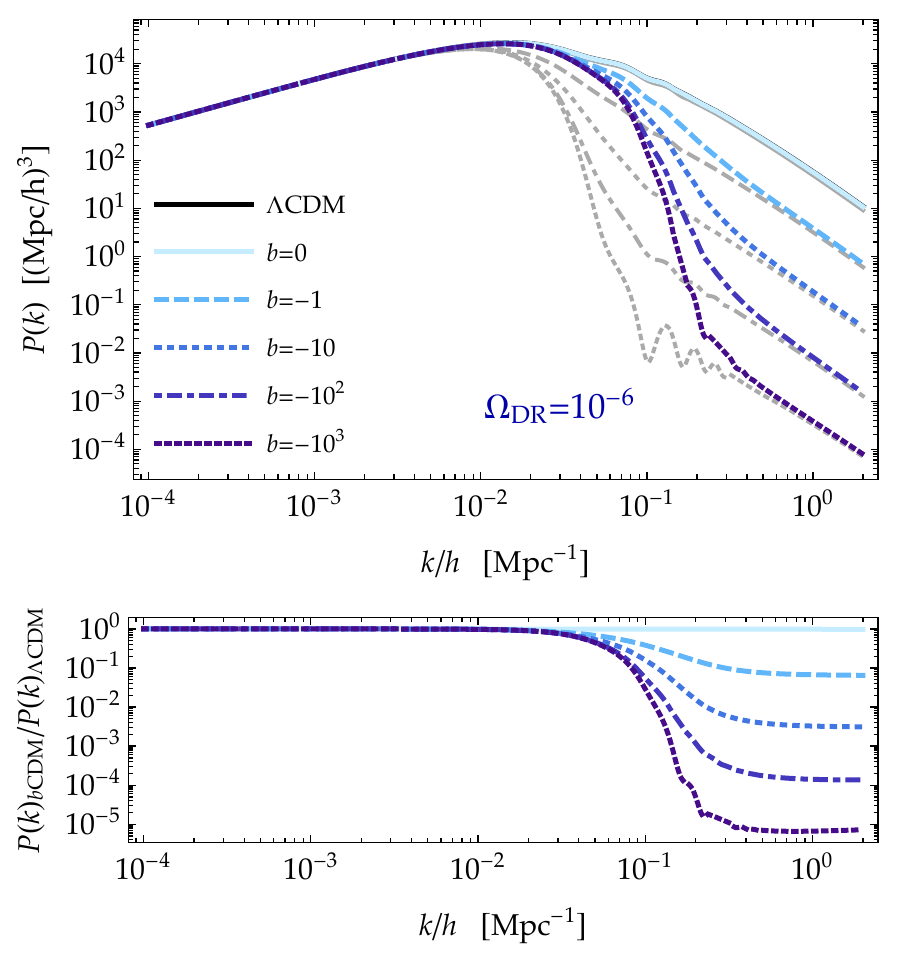}\\
\vspace{0.5cm}
\includegraphics[width=0.48\textwidth]{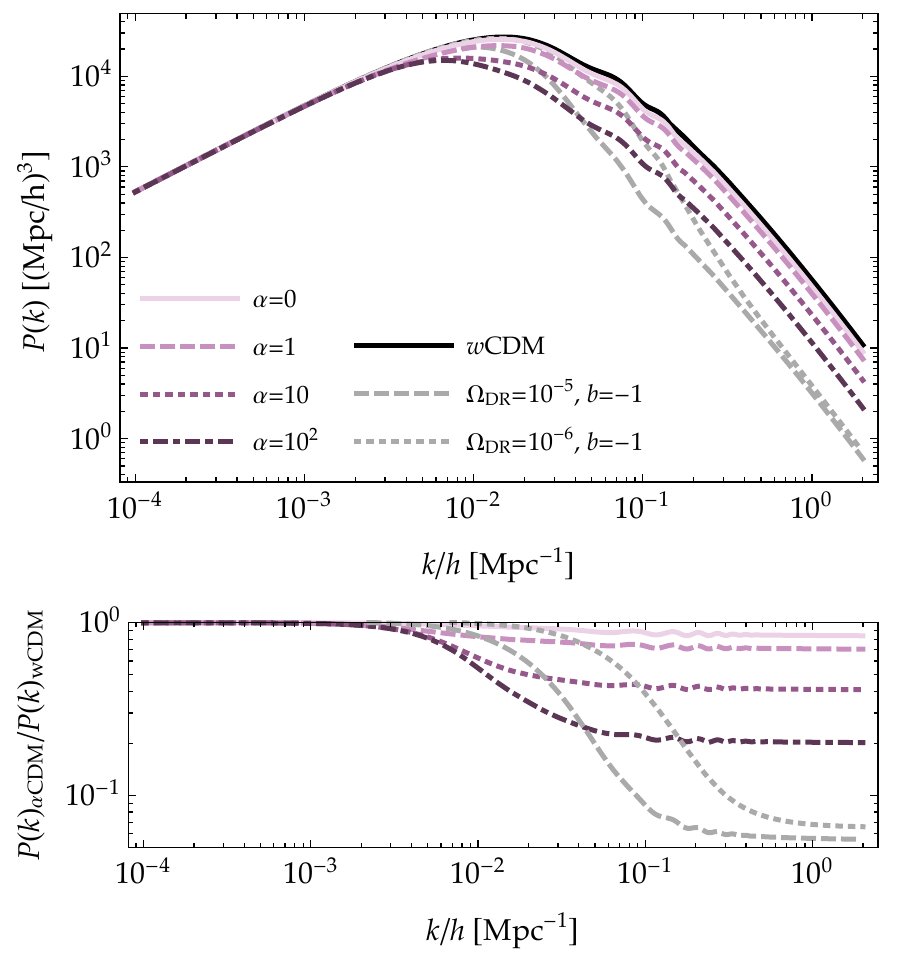}
\hspace{0.2cm}
\includegraphics[width=0.48\textwidth]{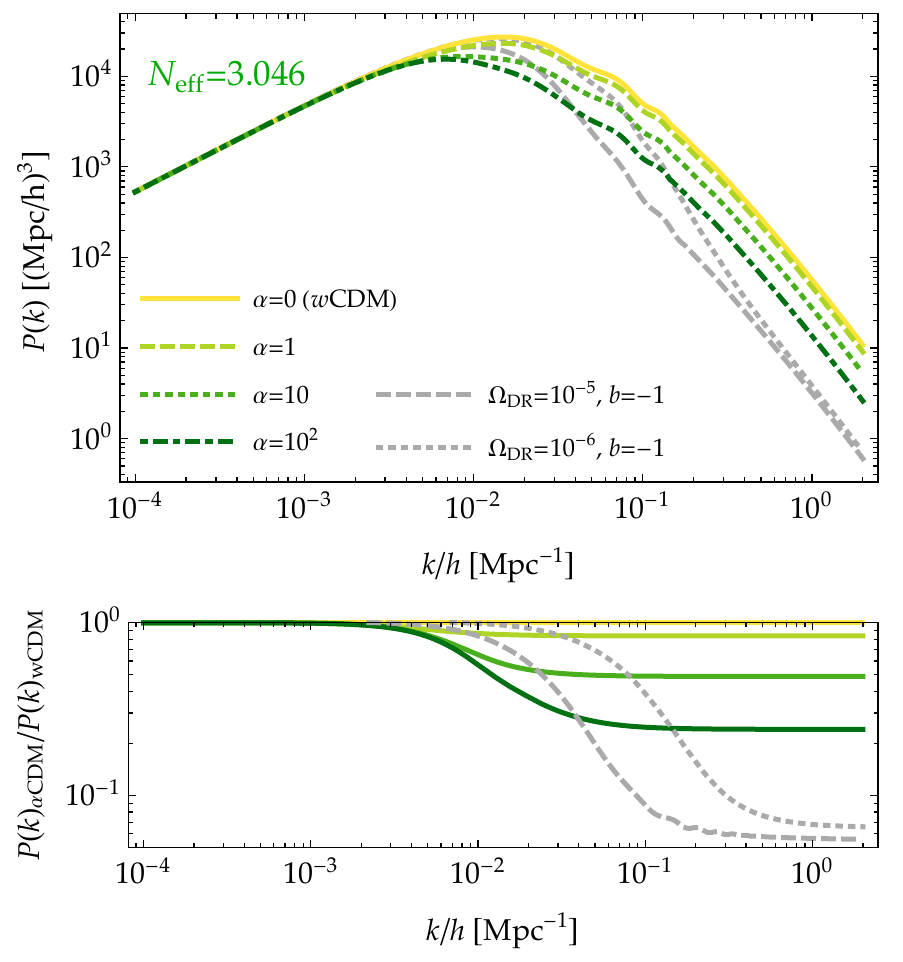}
\caption{
{\bf Upper panels:} The matter power spectra $P(k)$ 
in $\Lambda$CDM and $b$CDM models are shown, separately for 
$\Omega_{\rm DR}=10^{-5}$ and $\Omega_{\rm DR}=10^{-6}$ and 
different values of the interaction parameter $b$. 
In the left (right) panel, dashed grey lines indicate 
the $P(k)$ corresponding to 
$\Omega_{\rm DR}=10^{-6}$ ($\Omega_{\rm DR}=10^{-5}$) for better illustration of the effects of the two parameters. We also show the ratio of the power spectra relative to $\Lambda$CDM, where we can see how there is the suppression controlled by $b$ 
below a scale determined by $\Omega_{\rm DR}$. 
{\bf Lower panels:} The matter power spectrum $P(k)$ 
in $\alpha$CDM is displayed for different values of the interaction parameter $\alpha$. The case $\alpha=0$ corresponds to $w$CDM with $w=-0.98$. The $b$CDM models with $\Omega_{\rm DR}=10^{-5},10^{-6}$ 
and $b=-1$ are also included for comparison.} 
\label{fig:pkall}
\end{figure}

In Fig.~\ref{fig:clsDR}, the temperature and polarisation CMB angular power spectra are displayed in $b$CDM, together with $\Lambda$CDM, for $\Omega_{\rm DR}=10^{-5}$ and several values of the interacting parameter $b$. 
In all cases, only small scales are sensitive to the presence 
of the interaction, as expected. 
We have chosen the value $\Omega_{\rm DR}=10^{-5}$ to highlight the effects of the interacting model, but more reasonable values (compatible with observations) give a much milder effect. We can see that the peaks are substantially modified and this originates from the presence of a large dark radiation component at the recombination epoch so the main effect comes from $\Omega_{\rm DR}$. We also see however that the interaction parameter $b$ also contributes to the net effect on the peaks of the TT power spectrum. This is also reasonable since decoupling occurs in the matter dominated epoch, so there is a substantial amount of DM that will in turn affect the peaks via the interaction with the dark radiation, which also makes a fairly large contribution to the total radiation content of the universe for the chosen value of $\Omega_{\rm DR}$. As we will comment in more detail below, this large effect on the CMB peaks will represent an obstruction and will set a bound on the maximum suppression of clustering that we can have while being compatible with CMB data.

The angular power spectra of the $\alpha$CDM model are shown in Fig.~\ref{fig:clsEM} for some different values of the interaction parameter $\alpha$. In this case, the fiducial (non-interacting) model is $w$CDM, with $w=-0.98$. Furthermore, we have set 
$N_{\rm eff}=4.046$ to have a similar amount of extra dark radiation as in Fig.~\ref{fig:clsEM}. As a matter of fact, this high value of 
$N_{\rm eff}$ is entirely responsible for the effect on the acoustic peaks of the TT power spectrum and the interaction parameter $\alpha$ 
has no effect.
Unlike $b$CDM, the temperature power spectrum in $\alpha$CDM is actually most significantly modified at large scales. This is as expected because both the interaction and DE are relevant at very late times. For $b$CDM, however, the dark radiation component is most important at early times so the interaction plays an important role already at recombination time. This is one of the crucial differences between both scenarios. In both interacting scenarios, CDM interacts with a pressurefull fluid driven by the relative motion of both (thus exhibiting similar features), but such a fluid is most important at early times in $b$CDM model (dark radiation), while in $\alpha$CDM it becomes more relevant at low redshifts (as DE). This distinctive feature is crucial in explaining some of the differences between the two models we will find in our fit to data below. Finally, the polarisation power spectra are mainly modified at small scales for both models, showing an increase in the amplitude of high-$\ell$ oscillations that can be 
associated to lensing effects. 

The effects of the interactions on the matter power spectrum are shown in Fig.~\ref{fig:pkall}. The upper panels present the power spectra in $b$CDM for $\Omega_{\rm DR}= 10^{-5}$ (left) and $\Omega_{\rm DR}=10^{-6}$ (right) together with their ratios relative to $\Lambda$CDM. We can see the expected suppression of power on small scales that will be the main effect driving the alleviation of the $\sigma_8$ tension. 
The role of both parameters in this suppression is straightforward to analyse: $b$ controls the amount of suppression and $\Omega_{\rm DR}$ determines the largest scale that undergoes the suppressed matter 
power so that a higher value of $\Omega_{\rm DR}$ widens the range of suppressed scales. 
These results are in agreement with the analytic estimates and numerical results of the simplified scenario performed in Ref.~\cite{Jimenez:2020npm}. 
In the limiting case of small $\Omega_{\rm DR}$, all the suppressed scales fall into the non-linear regime. As commented above, this effect will represent an obstruction for the completely satisfactory resolution of the $\sigma_8$ tension because lowering $\sigma_8$ via the interaction requires large values of $\Omega_{\rm DR}$, 
but these large values will eventually conflict with the CMB (and eventually with BBN if dark radiation was present already at that time) because of the excess of dark radiation that, among other effects, greatly impacts the position and height of the first acoustic peaks (see Fig.~\ref{fig:clsDR}). 

In the lower panel of Fig.~\ref{fig:pkall}, we plot the matter power spectrum in the $\alpha$CDM model for different values of the interaction 
parameter $\alpha$. 
The case $\alpha= 0$ corresponds to $w$CDM model with $w =-0.98$. 
A much more thorough discussion about the evolution of 
perturbations and different effects is given 
in Ref.~\cite{Figueruelo:2021elm}, so we will not repeat it here. 
Let us simply note that the effects on the matter power spectrum are qualitatively similar to those in $b$CDM, i.e., the interaction drives the suppression of power on small scales. In this case, however, we do not have the tension between lowering the value of $\sigma_8$ and an excess of radiation at high redshifts. 
This will explain why the $\alpha$CDM model permits smaller values of $\sigma_8$ while being compatible with CMB data, thus showing a slightly better agreement with cluster counts and 
weak lensing data in the 
$\Omega_m$ - $\sigma_8$ plane. 
Now that we have introduced the models and gained some intuition on the most relevant effects, let us turn to the confrontation to observations.

\section{Observational constraints}

The preceding sections have been devoted to introducing the class of models under consideration in this work and analysing the evolution of the perturbations within this scenario together with their effects on the main cosmological observables. We will now proceed to perform the corresponding fit to observational data to test our expectations regarding the cosmological tensions and set precise limits on the interaction parameters. 
For this purpose, we have made use of the Markov-Chain Monte-Carlo 
codes CosmoMC~\cite{Lewis:2002ah,Lewis:2013hha} and Montepython~\cite{Brinckmann:2018cvx,Audren:2012wb}. The obtained chains, which are considered as converged when the Gelman-Rubin test satisfies $\vert R-1\vert < 0.01$, have then been analysed with GetDist~\cite{Lewis:2019xzd}. 
In the following, we use the notations $\Omega_b$, $\Omega_c$, 
and $\Omega_m=\Omega_b+\Omega_c$ to describe today's density 
parameters of baryons, CDM, and total non-relativistic matter, respectively.

In all cases, we vary all the standard (cosmological and nuisance) parameters with usual flat priors in addition to the new parameters of the corresponding interacting model. The considered primary cosmological parameters are the baryon density as $100\Omega_b h^2$, the CDM density as $\Omega_c h^2$, the scalar spectral index $n_s$, the amplitude of primordial perturbations as $10^9 A_s$, the optical depth $\tau_{\rm reio}$ and the angle corresponding to the comoving sound horizon at recombination as $100\theta_s$, 
with today's Hubble parameter $H_{0}=100h \Hunits$.
For the interacting parameters $b$ and $\alpha$, we set flat logarithmic priors with appropriate limits to avoid any bias on the final distributions. For stability reasons, we restrict to positive values of $\alpha$ \cite{Figueruelo:2021elm} and negative values of $b$ \cite{Jimenez:2020npm}. One concern might be that the logarithmic prior does not allow to strictly recover the non-interacting case, but, in practice, a sufficiently small value of the interaction parameters is indistinguishable from the interaction-less situation. We have thus chosen flat priors on $\log_{10}\vert b\vert\in[-4,0]$ and $\log_{10}\vert \alpha\vert\in[-4,4]$ for $b$CDM and $\alpha$CDM, respectively. The logarithmic priors are chosen to improve the convergence of the chains as well as the possibility of exploring more thoroughly several orders of magnitude. We have checked that varying the limits and/or using flat priors on these parameters do not significantly alter the final results (see also Ref.~\cite{Figueruelo:2021elm} where this point is illustrated in more detail for the $\alpha$CDM model). The priors on the parameters governing the amount of extra radiation ($\Omega_{\rm DR}$ and $N_{\rm eff}$ for $b$CDM and $\alpha$CDM, respectively) are also flat with appropriate boundaries, so that the dark radiation never dominates. In practice, we take $\Omega_{\rm DR}\in[10^{-8},10^{-5}]$ for $b$CDM and $N_{\rm eff}\in [0,10]$ for $\alpha$CDM. When varying $N_{\rm eff}$ we start from the neutrino sector of the baseline scenario used in Planck2018 \cite{Aghanim:2018eyx} with $N_{\rm eff}=3.046$ corresponding to two massless neutrinos and one massive neutrino of mass $0.06$~eV and attribute the difference $\Delta N_{\rm eff}=N_{\rm eff}-3.046$ to the massless sector. Finally, the $\alpha$CDM model also has DE equation of state $w$ as a free parameter and we follow Refs.~\cite{Asghari:2019qld,Figueruelo:2021elm} to choose a flat prior over $w>-1$. We do not go to the phantom region $w<-1$ in order to avoid the presence of instabilities (see Ref.~\cite{Figueruelo:2021elm} for more details). We will also give the constraints for the following derived parameters: the reionisation redshift  $z_{\rm reio}$, today's Hubble expansion rate $H_0$, the amplitude of matter perturbations $\sigma_8$, and the total non-relativistic matter $\Omega_m$.

\subsection{Constraints from CMB+BAO+SnIa}

\begin{figure}[!t] 
\includegraphics[width=0.9\textwidth]{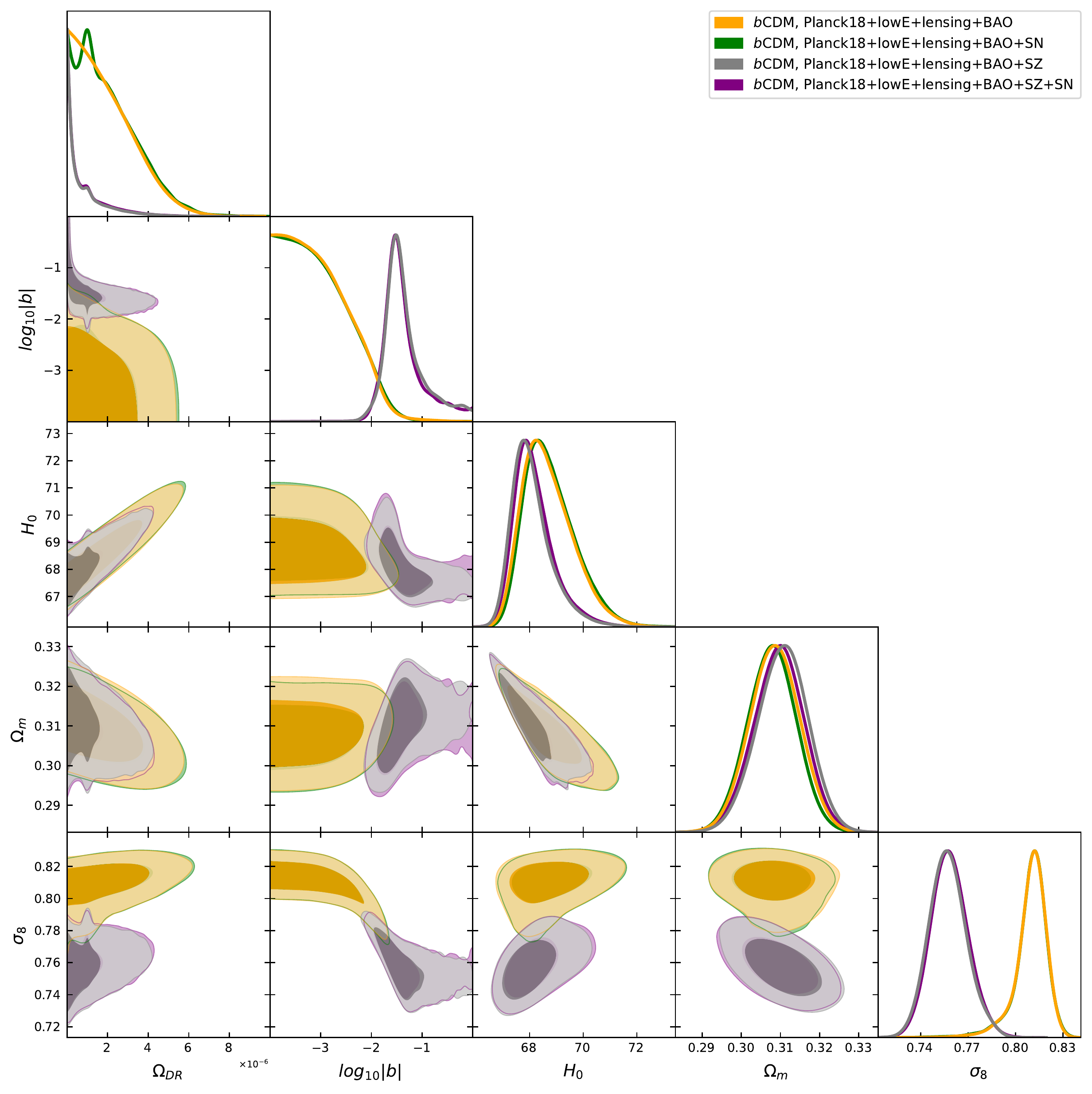}
\caption{Observational constraints on the $b$CDM model 
when using Planck2018 and BAO data (yellow), 
Planck2018, BAO and SnIa data (green), 
Planck2018, BAO and Planck SZ data (grey) and all the previous data sets (purple). 
We see how including the Planck SZ data is crucial 
to constrain the interaction parameter $b$. 
With the SZ data, the contours 
in the $\sigma_8$ - $\Omega_m$ plane is also shifted 
towards the region with smaller values of $\sigma_8$.
}
\label{Fig:triangleDR}
\end{figure}

\begin{figure}[!t]
\includegraphics[width=0.9\textwidth]{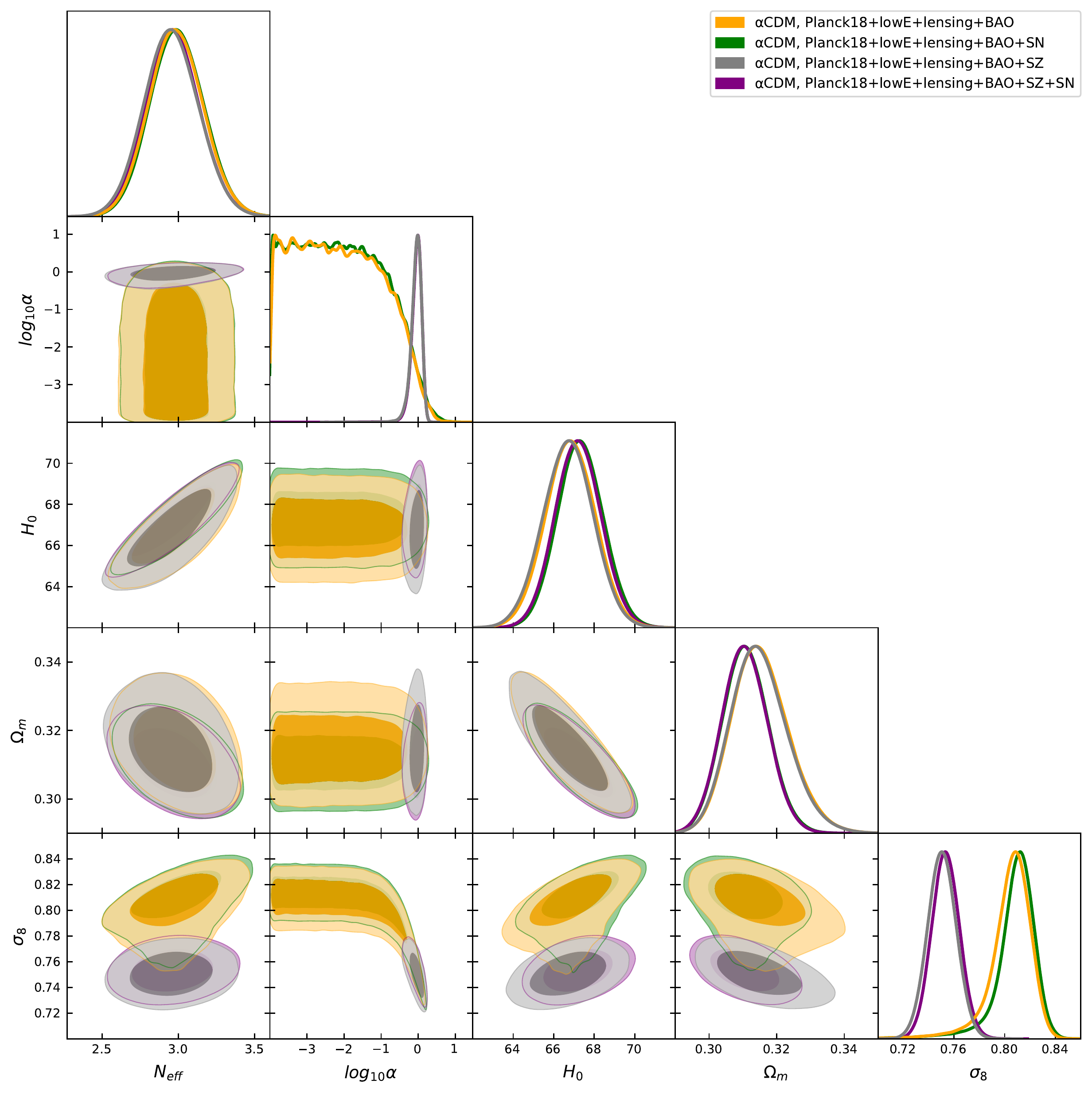}
\caption{Observational constraints on the $\alpha$CDM model
when using Planck2018 and BAO data (yellow), 
Planck2018, BAO and SnIa data (green), 
Planck2018, BAO and Planck SZ data (grey) and all the previous data sets (purple). The inclusion of Planck SZ data is crucial to constrain the coupling parameter $\alpha$ and to lower the value of $\sigma_8$ without affecting $\Omega_m$.} 
\label{Fig:FitElastic_triangle}
\end{figure}

We will first perform an analysis with the most standard data sets given by Planck2018 containing high-$\ell$ and low-$\ell$ from CMB temperature (TT), polarisation (EE), cross correlation of temperature and polarisation (TE) and the CMB lensing power spectrum~\cite{Aghanim:2018eyx,Aghanim:2019ame}, BAO combined data~\cite{Alam:2016hwk,2011MNRAS.416.3017B,Ross:2014qpa}, 
and SnIa data from Pantheon~\cite{Scolnic:2017caz}. 
From this analysis, we will obtain a more precise understanding on the performance of the models in alleviating the cosmological tensions with respect to the $\Lambda$CDM model. We will see that the tensions are indeed reduced, but some discomfort is still present when attempting to alleviate both the $H_0$ and $\sigma_8$ tensions simultaneously. We will not include local measurements of $H_0$ because they are in tension with the Planck2018 data and our goal is to explore the possibility of alleviating this tension. Including these measurements, besides being somewhat incorrect due to the existing incompatibilities, could give a fake alleviation of the tension precisely driven by the higher value of local measurements. From our analysis without the local measurements of $H_0$, 
we will unveil to what extent the presence of the extra dark radiation permits higher values of $H_0$ in view of the considered datasets (Planck2018+BAO+SnIa) and, thus, how much the tension can be alleviated. Since the SH0ES \cite{Riess:2019cxk} measurement provides the value with the highest tension, we will use it as a reference in our analysis below. For the same reasons, we will not include for the moment other low-redshift probes such as cosmic shear or cluster counts, although we will come back to these in the next section. 
Let us then discuss the results for each model separately:

\begin{itemize}
  
\item {\bf $b$CDM:} The 2-dimensional constraints on the model parameters in $b$CDM are shown in Fig.~\ref{Fig:triangleDR} and the 1-dimensional 
limits are given in Table~\ref{tab:bestfit}. 
We have considered different combinations of datasets in order to analyse their compatibility and individual effects on the values of parameters. 
The $H_0$ - $\Omega_{\rm DR}$ plane in Fig.~\ref{Fig:triangleDR} shows that higher values of $H_0$ are allowed by Planck2018+BAO+SnIa data thanks to the presence of the dark radiation component, as expected. The Hubble constant is constrained to be
\be
H_0=68.79^{+0.67+1.92}_{-1.13-1.63}
~\,{\rm km}~{\rm s}^{-1}
~{\rm Mpc}^{-1}\,,
\label{H0bo1}
\ee
where the first and second values in upper and 
lower indices represent $1\sigma$ and $2\sigma$ errors, respectively. 
The upper $2\sigma$ limit is then 
$70.71 {\rm km}~{\rm sec}^{-1}~{\rm Mpc}^{-1}$ as compared to $68.48{\rm km}~{\rm sec}^{-1}
~{\rm Mpc}^{-1}$ in $\Lambda$CDM. 
We thus obtain a substantial reduction of the tension with the SH0ES \cite{Riess:2019cxk} measurement which is driven by the extra dark radiation. For $\sigma_8$ we obtain 
\be
\sigma_8=0.810^{+0.010+0.019}_{-0.006-0.020}\,.
\ee
As we see in Table~\ref{tab:bestfit}, the mean value 
is similar to the constraint on $\sigma_8$ for $\Lambda$CDM. However, the $2\sigma$ limit is considerably broader for the interacting 
model (roughly a factor of 2), so the interacting scenario permits lower values of $\sigma_8$. 
In this case, the  
$\sigma_8$ - $\log_{10}\vert b\vert$ plane shows 
how the interaction is responsible for the allowed lower values of $\sigma_8$. The interaction parameter is only constrained with the upper bound
\be
\log_{10}\vert b\vert<-3.1 ^{+ 0.3} {}^{ +1.1}.
\ee

Finally, let us notice how the $\sigma_8$ - $H_0$ plane in Fig.~\ref{Fig:triangleDR} clearly illustrates the orthogonality of alleviating both tensions simultaneously since in regions with 
lower values of $\sigma_8$ {\it or} higher values of $H_0$ (with respect to $\Lambda$CDM) are allowed by the data, but the sweet region in that plane with higher $H_0$ {\it and} lower $\sigma_8$ is outside the $2\sigma$ region. 

\item {\bf $\alpha$CDM:} 
The results for the $\alpha$CDM model are shown in Fig.~\ref{Fig:FitElastic_triangle} 
and in Table~\ref{tab:bestfit}. We observe that the observational constraints are rather similar to those of the $b$CDM model with $\alpha$ playing the role of $b$ and $N_{\rm eff}$ giving similar effects to $\Omega_{\rm DR}$. A difference is that we allow $N_{\rm eff}$ to explore values smaller than $3.046$ that would correspond to a negative $\Omega_{\rm DR}$, but, for obvious reasons, we always keep $\Omega_{\rm DR}$ positive. All the relevant 2-dimensional planes exhibit similar constraints for the cosmological parameters. 
In the $H_0$ - $N_{\rm eff}$ plane, we see again how the varying 
$N_{\rm eff}$ allows higher values of $H_0$. This behaviour is consistent with the 
Planck2018 results \cite{Aghanim:2018eyx} 
and other previous studies for scenarios with varying 
$N_{\rm eff}$~\cite{Ballesteros:2020sik,Seto:2021xua}.\footnote{The idea behind these scenarios is to take advantage 
of the known $N_{\rm eff}$ - $H_0$ degeneracy without spoiling the CMB that breaks this degeneracy 
(see e.g., \cite{Bashinsky:2003tk,2013PhRvD..87h3008H}). 
Let us also emphasise that we do not include the local measurement of $H_0$ as discussed at the beginning of this section. This is different from other studies in the literature where the SH0ES measurement is included, 
in which case it naturally drives the obtained result for $H_0$ to higher values. This cannot be interpreted as an alleviation of the Hubble tension by itself because it is precisely a consequence of combining datasets 
that are in tension.} 
The Hubble constant for $\alpha$CDM is
\be
H_0=67.31^{+1.13+2.24}_{-1.12-2.19}
~\,{\rm km}~{\rm s}^{-1}
~{\rm Mpc}^{-1}\,,
\ee
which shows that the tension with the local measurements is still alleviated but to a less extent than $b$CDM. Let us recall in this respect that varying $N_{\rm eff}$ has additional constraints from higher-order moments that we exclude in $b$CDM when we assume $\Omega_{\rm DR}$ 
to be a perfect fluid. 
On the other hand, the value of $\sigma_8$ 
is found to be
\be
\sigma_8=0.808^{+0.017+0.031}_{-0.008-0.035}\,,
\ee
whose mean value is again similar to that of $\Lambda$CDM, but whose $2\sigma$ limits are broader and, in turn, permit slightly lower values of $\sigma_8$ as compared to $b$CDM. Thus, while $b$CDM performs slightly better regarding the $H_0$ tension, 
$\alpha$CDM allows lower values of $\sigma_8$. 
As we observe in 
Fig.~\ref{Fig:FitElastic_triangle}, the 
$\sigma_8$ - $H_0$ plane shows a certain orthogonality to ameliorate both tensions simultaneously. 
Like the $b$CDM model, the interaction parameter 
$\alpha$ is only constrained from above as
\be
 \log_{10} \alpha<-2.1^{+0.7+1.9}\,.
\ee
\end{itemize}

\begin{figure}[ht!]
\includegraphics[width=0.32\textwidth]{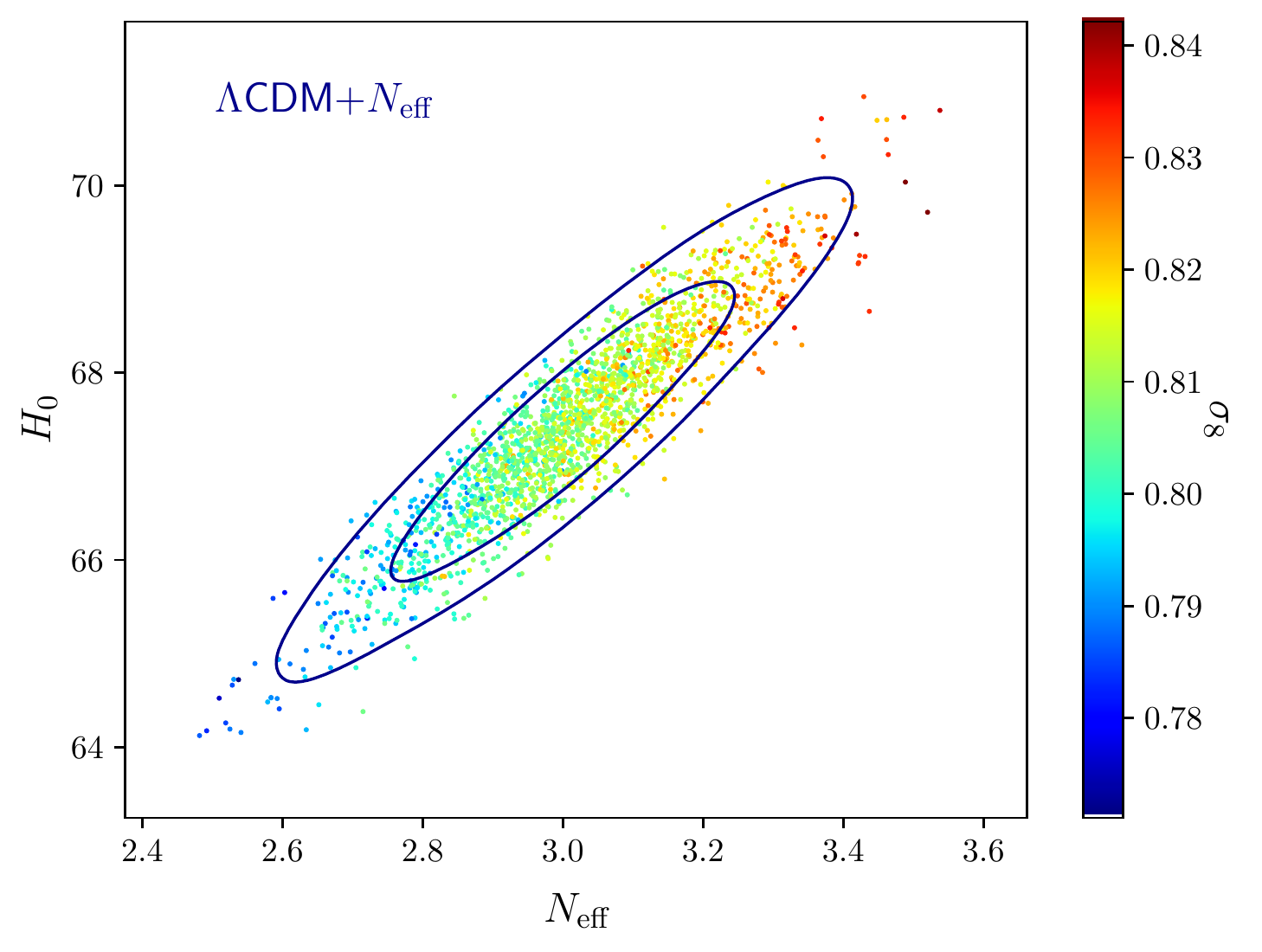}
\includegraphics[width=0.32\textwidth]{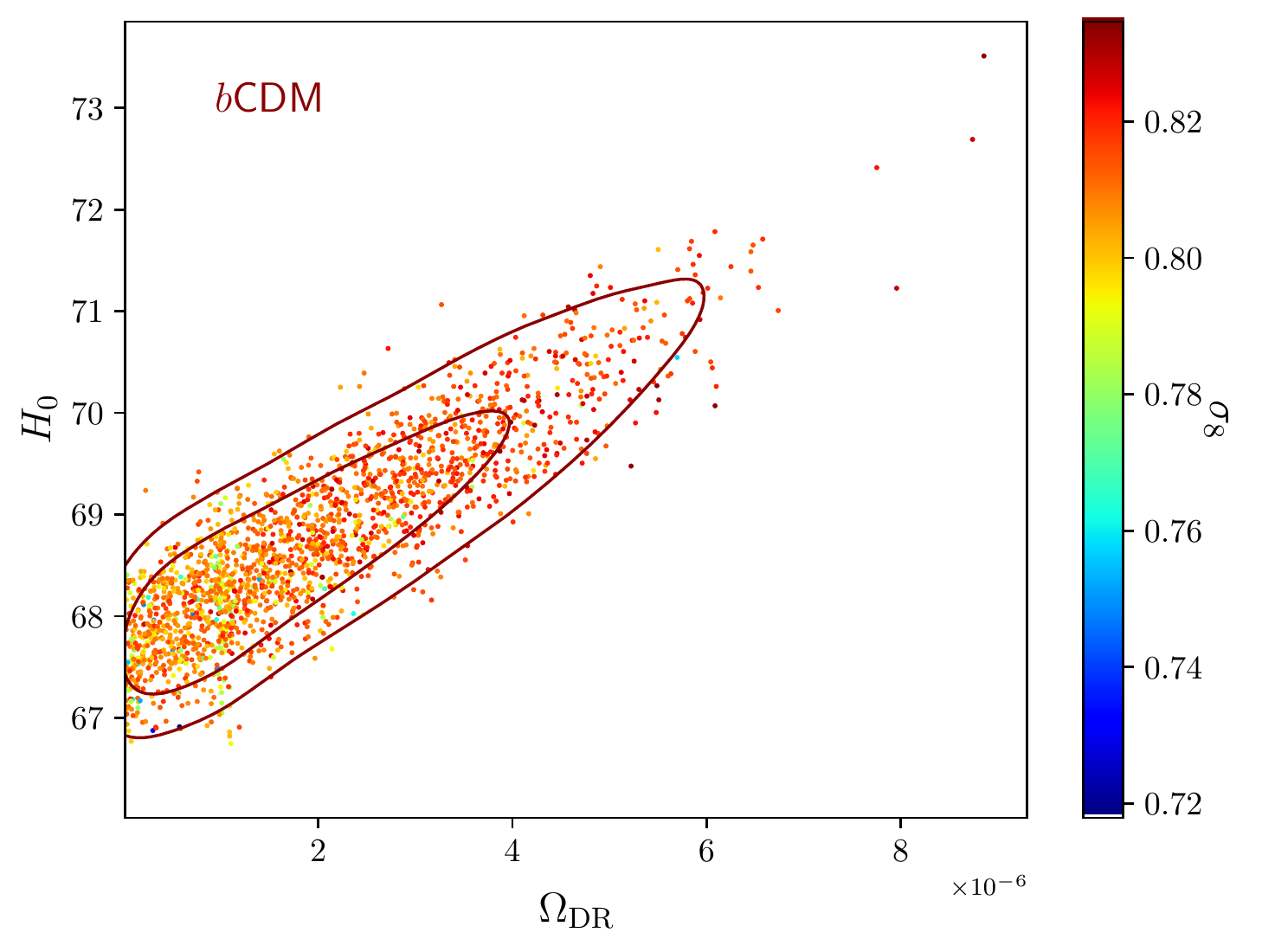}
\includegraphics[width=0.32\textwidth]{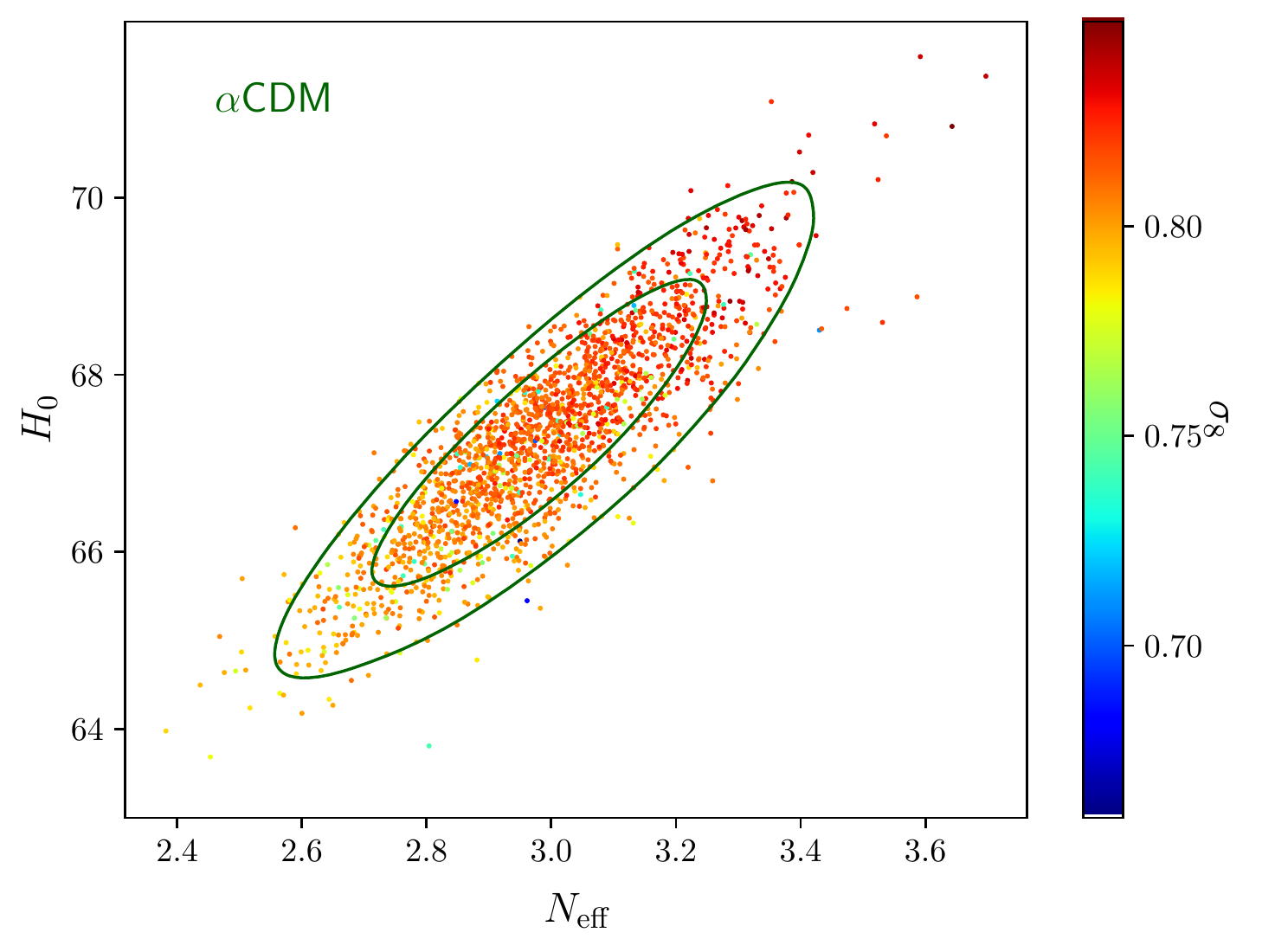}
\caption{In this Figure we show the 2-dimensional posteriors in the plane of $H_0$ and the abundance of extra radiation determined by $N_{\rm eff}$ or $\Omega_{\rm DR}$ together with samples from the Planck2018+BAO+SN chains coloured by the value of $\sigma_8$. As expected, we corroborate the correlation between $H_0$ and the amount of extra radiation so that increasing the dark radiation abundance leads to higher values of $H_0$. In the left panel, we can see the additional correlation with $\sigma_8$ in $\Lambda$CDM$+N_{\rm eff}$ so that increasing $H_0$ at the expense of 
dark radiation abundance is tightly linked to an increase of $\sigma_8$, i.e., to a worsening of the $\sigma_8$ tension. In the middle and right panels, it is apparent how, while the correlation between $H_0$ and the amount of dark radiation is maintained, the presence of the interaction removes the correlation with $\sigma_8$ so that an alleviation of the $H_0$ tension does not incur into a worsening of the $\sigma_8$ tension. For the $b$CDM model there is a very mild correlation because the suppression of structures is due to an interaction with the dark radiation, so the more dark radiation the larger the suppression. However, the suppression is mainly driven by the interaction parameter $b$, as clearly seen in Fig.~\ref{Fig:oms8interactions}. In $\alpha$CDM, since the interaction is with the DE component, the correlation between higher $H_0$ and higher $\sigma_8$ completely disappears.}
\label{Fig:H0Neffsigma} 
\end{figure}

\begin{figure}[ht!]
\includegraphics[width=0.48\textwidth]{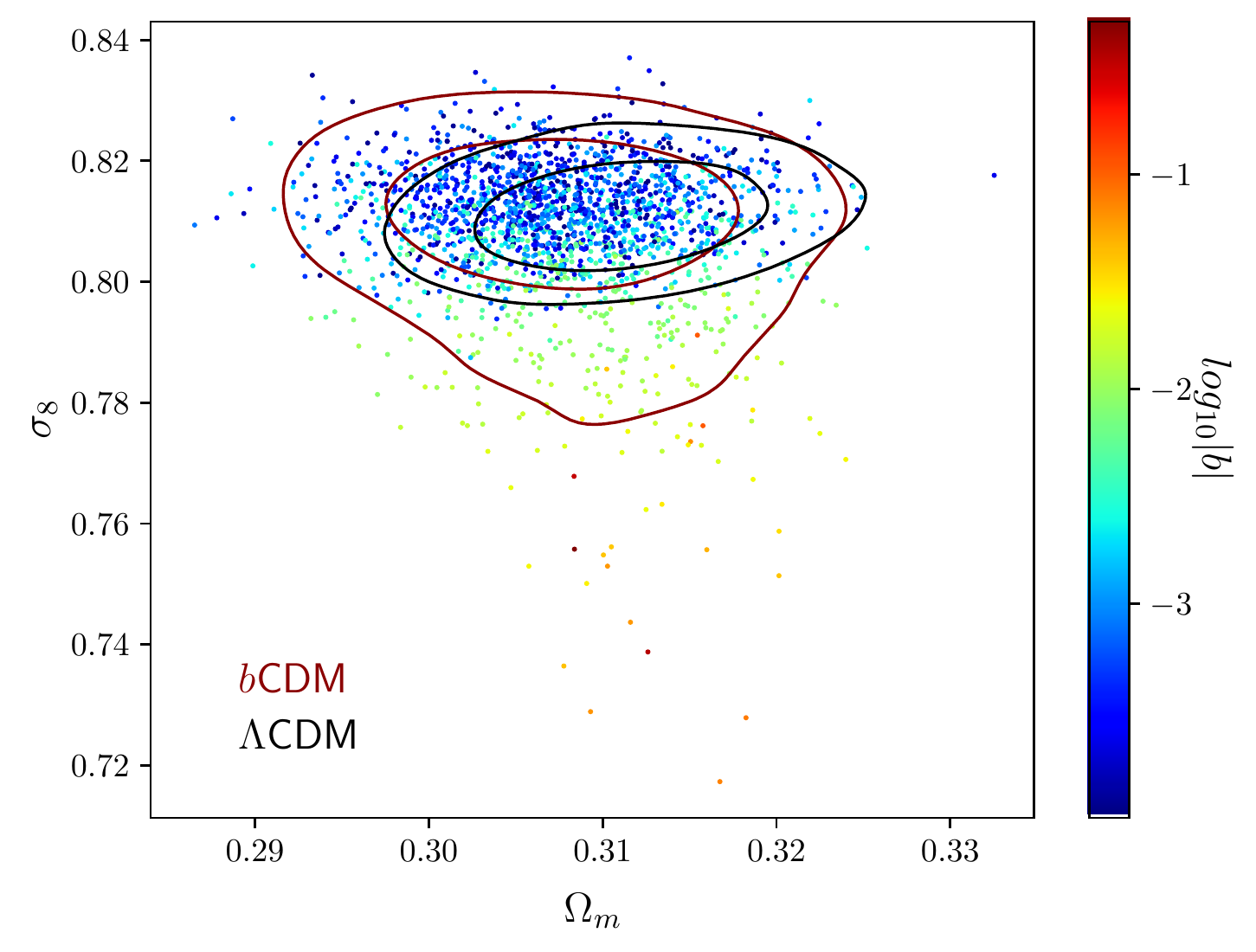}
\includegraphics[width=0.48\textwidth]{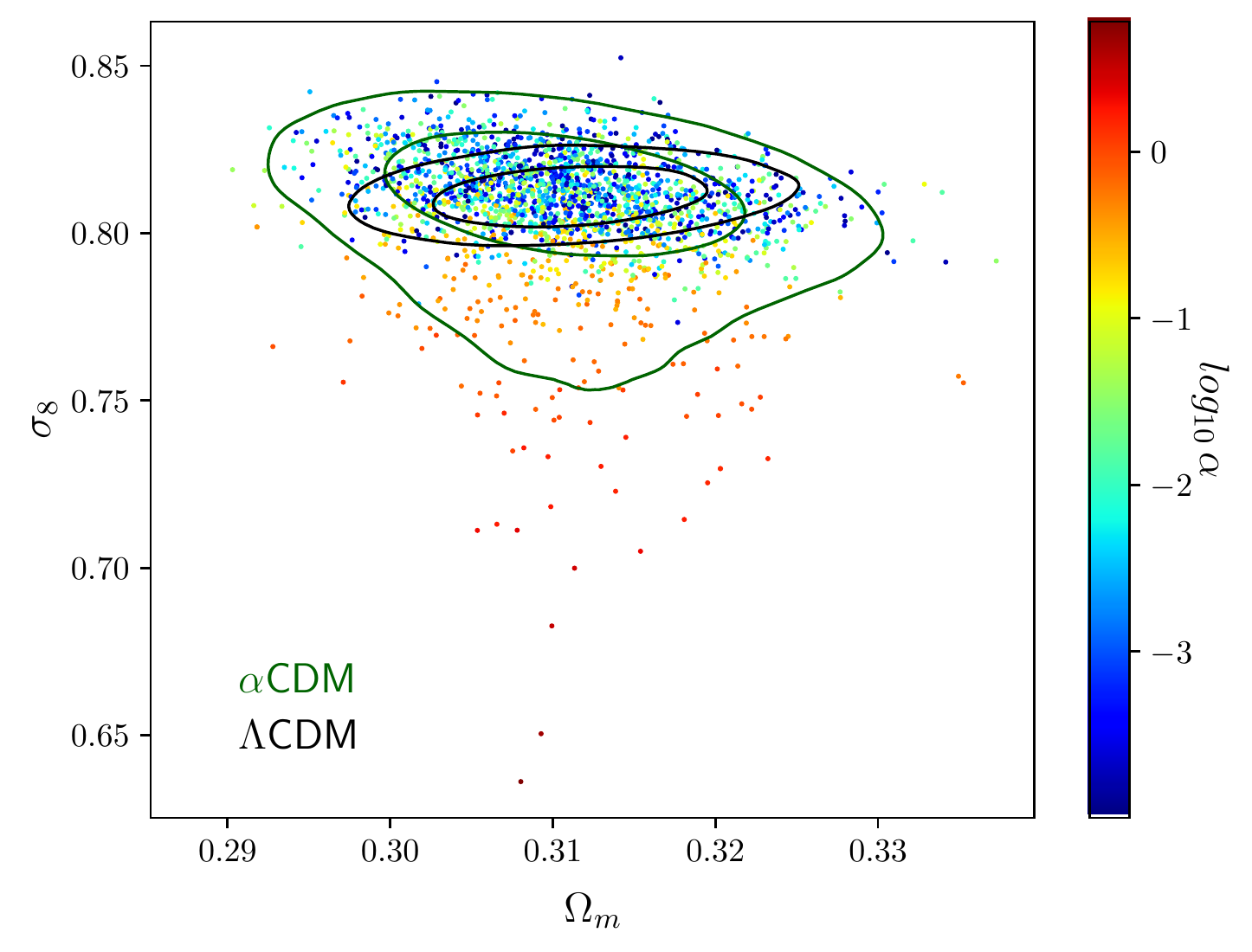}
\caption{In this Figure we show samples of the chains for Planck2018+BAO+SN colour-coded according to the interaction parameters. The solid lines correspond to the 2-dimensional posterior distributions for the corresponding models. In both cases, we can see how the distributions spread towards lower values of $\sigma_8$ than in $\Lambda$CDM without a substantial effect on $\Omega_m$. Furthermore, it is clear that the samples in the lower part of the distributions correspond to higher values of the interaction parameters. } 
\label{Fig:oms8interactions}
\end{figure}

In summary, we have corroborated how the presence of the extra radiation in both models permits to alleviate the $H_0$ tension, while the interaction, that suppresses the clustering, allows to have lower 
values of $\sigma_8$. 
However, the region with higher $H_0$ and lower $\sigma_8$ than in $\Lambda$CDM is not favoured for neither of the interacting models, so that alleviating both tensions simultaneously 
seems to be difficult. Let us notice nonetheless that improving one of the tensions does not incur into a worsening of the other. In other words, the existing correlation between $H_0$ and $\sigma_8$ in $\Lambda$CDM and other scenarios is not present in our elastic interacting models. This can be clearly appreciated from the $\sigma_8$ - $H_0$ plane in Fig.~\ref{Fig:triangleDR} for the $b$CDM model. For $\alpha$CDM, we can see some correlation in Fig.~\ref{Fig:FitElastic_triangle} but still less pronounced than in $\Lambda$CDM. 
In Fig.~\ref{Fig:H0Neffsigma} we can clearly see how the extra radiation allows to increase the value of $H_0$ without necessarily increasing $\sigma_8$, which does not happen even in the extended $\Lambda$CDM$+N_{\rm eff}$ model so it is owed to the crucial presence of the interactions. This is also illustrated in Fig.~\ref{Fig:oms8interactions} where we can see that lower values of $\sigma_8$ are allowed for the interacting models as compared to $\Lambda$CDM and this region precisely corresponds to having larger values of the coupling parameters.

\begin{table}
\begin{center}
\renewcommand{\arraystretch}{1.8}
\begin{tabularx}{0.8\textwidth}{ |c|| *{3}{Y|} }
	\hline
	\hline
	 \multicolumn{4}{|c|}{ Data: Planck18+BAO+SN }\\ 
	\hline
	  Parameter & {\bf $\Lambda$CDM  }  &  {\bf $b$CDM }  & {\bf $\alpha$CDM}\\ 
	\hline 
	\hline 
 $100\Omega_{b} h^2$ &  $2.242 _{-0.013}^{+0.013} {}_{-0.027}^{+0.026}$  &  $2.26 _{-0.018}^{+0.015} {}_{-0.031}^{+0.034}$ &  $ 2.239_{-0.019}^{+0.019} {}_{-0.036}^{+0.037}   $ \\\hline
 $\Omega_{c}h^2$     &  $0.1193_{- 0.0009}^{+ 0.0009} {}_{-0.0018}^{+0.0018}  $  &  $0.1224  _{- 0.0029}^{+ 0.0017} {}_{ - 0.0041}^{ + 0.0049}  $   &  $0.1184_{-0.0029}^{+0.0029} {}_{-0.0056}^{+0.0058}$   \\ \hline 
 $n_{s}$              &   $ 0.966 _{-0.004}^{+0.004} {}_{-0.007}^{+0.007} $ & $0.967 _{- 0.004} ^{+ 0.004} {}_{- 0.007} ^{+ 0.007} $ &  $0.965_{-0.007}^{+0.007} {}_{-0.014}^{+0.014}$  \\ \hline 
 $10^{9}A_{s}$        & $2.11_{-0.03}^{+0.03} {}_{- 0.06}^{+ 0.06} $   &  $2.11_{-0.03}^{+0.03} {}_{-0.06} ^{+0.06} $  &  $2.10_{-0.04}^{+0.03} {}_{-0.07}^{+0.07}$  \\ \hline  
 $\tau_{\rm reio}$       &  $0.057_{- 0.008}^{+ 0.007} {}_{-0.014}^{+0.015}  $   & $0.058 _{-0.008}^{+0.007} {}_{- 0.014}^{+0.015} $  &  $0.056_{-0.008}^{+0.007} {}_{-0.014}^{+0.015}$ \\ \hline 
 $w$                  &  $-1$  &  $-1$  &  $-0.980_{-0.019}^{+0.005} {}_{-0.020}^{+0.032} $  \\ \hline 
 $100~\theta_{s}$   &  $1.0410_{- 0.0003}^{+0.0003}  {}_{-0.0006}^{+0.0006} $  &  $1.0412 _{- 0.0003}^{+0.0003} {}_{-0.0006}^{+ 0.0006} $  &    $1.0421_{-0.0005}^{+0.0005} {}_{-0.0010}^{+0.0010}$    \\ \hline 
 \hline
 $z_{\rm reio}$        & $7.95_{- 0.71}^{+0.71} {}_{-1.43}^{+1.43} $  & $8.07 _{-0.72}^{+0.72} {}_{- 1.43}^{+ 1.46} $   &     $7.82_{-0.71}^{+0.72} {}_{-1.44}^{+1.45}$    \\ \hline 
 $H_0$ [km/s/Mpc]  &  $67.67 _{- 0.41}^{+0.41} {}_{-0.81}^{+0.81}$  & $ 68.79 _{-1.13}^{+0.67} {}_{- 1.63}^{+1.92} $ &     $67.31_{-1.12}^{+1.13} {}_{-2.19}^{+2.24}$    \\ \hline 
 $\sigma_8$  &   $0.811_{- 0.006}^{+0.006} {}_{-0.012}^{+0.012}$ & $ 0.810_{-0.006}^{+0.01} {}_{-0.02}^{+0.019}$  &  $0.808_{-0.008}^{+0.017} {}_{-0.035}^{+0.031} $   \\ \hline 
 $\Omega_{m }$  &  $0.311_{-0.006}^{+0.006} {}_{-0.011}^{+0.011} $  &  $0.308 _{-0.006}^{+0.006} {}_{- 0.012}^{+0.012} $  &  $0.311_{-0.007}^{+0.007} {}_{-0.013}^{+0.014} $    \\ \hline 
 $S_8=\sigma_8(\Omega_m/0.3)^{0.5}$ &  $ 0.826_{-0.011}^{+ 0.011} {}_{- 0.021}^{+0.020}$  &  $0.821 _{- 0.010}^{+0.012} {}_{-0.024}^{+ 0.024}$  &    $0.822_{-0.009}^{+0.018} {}_{-0.036}^{+0.032}$ \\ \hline 
 \hline
 $10^{6}\Omegadr$  &  $-$  & $<2.05^{+0.05} {}^{+2.71}  $   &  $-$    \\ 
 \hline
 $ N_{\rm eff}$  &  $3.046$  & $-$   &  $2.99_{-0.18}^{+0.18} {}_{-0.34}^{+0.35}$    \\ 
 \hline
 $ \log_{10} \vert b\vert$/$ \log_{10} \alpha$  &  $-$  &  $<-3.1 ^{+ 0.3} {}^{ +1.1}  $  &  $<-2.1^{+0.7} {}^{+1.9}$   \\ 
 \hline
 \hline
\end{tabularx} 
\end{center}
\caption{Mean likelihood values and $1\sigma$ and $2\sigma$ limits for the cosmological and derived parameters in the $\Lambda$CDM, 
$b$CDM, and $\alpha$CDM models. 
Here, $n_s$ and $A_s$ are the scalar spectral index and amplitude
of primordial curvatures perturbations, respectively, 
$\tau_{\rm reio}$ is the optical depth, $\theta_s$ is the angle corresponding to the comoving sound horizon at recombination, and $z_{\rm reio}$ is the redshift at reionization.
The corresponding observational contours for the two later models are shown in Fig.~\ref{Fig:triangleDR} and \ref{Fig:FitElastic_triangle}, respectively.}
\label{tab:bestfit}
\end{table}

\begin{table}
\begin{center}
\renewcommand{\arraystretch}{1.8}
\begin{tabularx}{0.9\textwidth}{ |c|| *{4}{Y|} }
	\hline
	\hline
	   {\bf $\Lambda$CDM }    & P18+BAO  & P18+BAO+SN   & P18+BAO+SZ & P18+BAO+SZ+SN \\ 
	\hline 
	\hline 
 $H_0$ [km/s/Mpc]  & ${67.62}_{-0.42}^{+ 0.42}{}_{-0.85}^{+0.83}$ 
 & $67.67 _{- 0.41}^{+0.41} {}_{-0.81}^{+0.81}$  
 & ${68.58}_{-0.39}^{+0.39} {}_{-0.76}^{+0.76}$ 
 & ${68.58}_{-0.38}^{+ 0.38} {}_{-0.75}^{+0.75}$   
 \\ \hline 
 $\sigma_8$  &  ${0.811}_{-0.006}^{+0.006}{}_{-0.012}^{+0.012}$ 
 &$0.811_{- 0.006}^{+0.006} {}_{-0.012}^{+0.012}$  
 & ${0.795}_{-0.005}^{+0.005} {}_{-0.010}^{+0.01}$ 
 & ${0.795}_{-0.005}^{+ 0.005} {}_{-0.010}^{+0.01}$     
 \\ \hline 
 $\Omega_{m}$  &  ${0.312}_{-0.006}^{+0.006} {}_{-0.011}^{+0.012}$  
 &  $0.311_{-0.006}^{+0.006} {}_{-0.011}^{+0.011} $  
 & ${0.299}_{-0.005}^{+0.005} {}_{-0.01}^{+0.01}$  
 & ${0.299}_{-0.005}^{+0.005} {}_{-0.009}^{+0.01}$  
 \\ \hline 
 $S_8=\sigma_8(\Omega_m/0.3)^{0.5}$ &  ${0.827}_{-0.011}^{+0.011} {}_{-0.021}^{+0.021}$ 
 & $ 0.826_{-0.011}^{+ 0.011} {}_{- 0.021}^{+0.020}$ 
 & ${0.793}_{-0.008}^{+0.008} {}_{-0.016}^{+0.016}$  
 & ${0.793}_{-0.008}^{+ 0.008} {}_{-0.015}^{+0.015}$  
 \\ \hline 
 $\chi^2_{\rm best\;fit}$ & 2783 
 &  3817 
 & 2807 
 & 3842
 \\ \hline
\hline
	   {\bf$b$CDM }    & P18+BAO  & P18+BAO+SN   & P18+BAO+SZ & P18+BAO+SZ+SN \\ 
	\hline 
	\hline 
 $H_0$ [km/s/Mpc]  &  ${68.70}_{-1.12}^{+ 0.69} {}_{-1.64}^{+1.9} $ & $ 68.79 _{-1.13}^{+0.67} {}_{- 1.63}^{+1.92} $   &  ${68.09}_{-0.87}^{+0.42}{}_{-1.30}^{+1.68}$ &  ${68.14}_{-0.86}^{+0.45} {}_{-1.32}^{+1.67}$ \\ \hline 
 $\sigma_8$  &  ${0.810}_{-0.006}^{+0.01} {}_{-0.02}^{+ 0.018} $       & $ 0.810_{-0.006}^{+0.01} {}_{-0.02}^{+0.019}$    &  ${0.757}_{-0.011}^{+0.011}{}_{0.022}^{+0.023}$  &  ${0.758}_{-0.012}^{+0.011} {}_{-0.022}^{+ 0.024}$   \\ \hline 
 $\Omega_{m}$  & ${0.309}_{-0.006}^{+0.006} {}_{-0.012}^{+0.012}$  &  $0.308 _{-0.006}^{+0.006} {}_{- 0.012}^{+0.012} $ &  ${0.311}_{-0.006}^{+0.006}{}_{-0.013}^{+0.012}$  &  ${0.31}_{-0.006}^{+0.006} {}_{-0.013}^{+0.012}$   \\ \hline 
 $S_8=\sigma_8(\Omega_m/0.3)^{0.5}$ &  ${0.822}_{-0.011}^{+0.013} {}_{-0.025}^{+0.025}$   &  $0.821 _{- 0.010}^{+0.012} {}_{-0.024}^{+ 0.024}$   &  ${0.770}_{-0.010}^{+0.010} {}_{-0.02}^{+0.02}$  &  ${0.770}_{-0.01}^{+0.010} {}_{-0.02}^{+0.02}$ \\ \hline 
 $\log_{10}{|b|}$  & $<{-3.1}^{+0.3} {}^{+1.1}$    &$ <{-3.1}^{+0.3} {}^{+1.1}$   &${-1.3}_{-0.4}^{+ 0.2} {}_{-0.6}^{+1.1}$   &${-1.3}_{-0.4}^{+0.2} {}_{-0.6}^{+1.1}$ \\ \hline
$\chi^2_{\rm best\;fit}$ &  2785 &  3820&   2791&   3827\\ \hline
 $\Delta$AIC & $6 $ & $7 $ & $ -12$ & $ -11$ \\ \hline 
\hline 
\hline
	   {\bf $\alpha$CDM}   & P18+BAO  & P18+BAO+SN   & P18+BAO+SZ & P18+BAO+SZ+SN \\ 
	\hline
	\hline 
 $H_0$ [km/s/Mpc]                    & $66.82_{-1.21}^{+1.21}{}_{-2.39}^{+2.40}$ & $67.31_{-1.12}^{+1.13}{}_{-2.19}^{+2.24}$     & $  66.71_{-1.23}^{+1.22}{}_{-2.43}^{+2.40}$           &    $  67.20_{-1.12}^{+1.12}{}_{-2.19}^{+2.24}$      \\ \hline 
 $\sigma_8$                          &  $0.804_{-0.010}^{+0.018}{}_{-0.035}^{+0.031}$       & $0.808_{-0.008}^{+0.017}{}_{-0.035}^{+0.0309} $  & $  0.751_{-0.011}^{+0.011}{}_{-0.022}^{+0.022}$       &    $  0.753_{-0.011}^{+0.011}{}_{-0.021}^{+0.022}$      \\ \hline 
 $\Omega_{m }$                       &  $0.315_{-0.009}^{+0.007}{}_{-0.016}^{+0.017}$      &  $0.311_{-0.007}^{+0.007}{}_{-0.013}^{+0.014} $ & $  0.315_{-0.009}^{+0.007}{}_{-0.016}^{+0.017}$        &    $  0.311_{-0.007}^{+0.007}{}_{-0.013}^{+0.014}$      \\ \hline 
 $S_8=\sigma_8(\Omega_m/0.3)^{0.5}$ &  $0.824_{-0.010}^{+0.017}{}_{-0.035}^{+0.032}$    &  $0.822_{-0.009}^{+0.018}{}_{-0.036}^{+0.032}$ & $  0.769_{-0.011}^{+0.011}{}_{-0.021}^{+0.021}$  &    $  0.766_{-0.010}^{+0.010}{}_{-0.020}^{+0.021}$        \\ \hline 
 $ \log_{10} \alpha$                      &  $<-2.1^{+0.7}{}^{+1.9}$          & $<-2.1^{+0.7}{}^{+1.9}$     & $  -0.06_{-0.10}^{+0.16}{}_{-0.29}^{+0.26}$    &    $  -0.05_{-0.10}^{+0.16}{}_{-0.28}^{+0.26}$       \\ \hline
 $\chi^2_{\rm best\;fit}$ &  2781 &  3817  &  2782 &  3818  \\ \hline 
  $\Delta$AIC & $ 4$ & $6 $ & $ -19$ & $ -18$ \\ \hline 
\hline 
\end{tabularx}
\end{center}
\caption{Mean likelihood values and $1\sigma$ and $2\sigma$ limits for some relevant derived parameters using different data sets, for the $\Lambda$CDM, 
$b$CDM, and $\alpha$CDM models. 
The corresponding observational contours for the two later 
models are shown in Fig.~\ref{Fig:triangleDR} and \ref{Fig:FitElastic_triangle}, respectively. We also provide the corresponding best-fit $\chi^2_{\rm best\;fit}\equiv -2 {\rm ln}(\mathcal{L}_{\rm max})$ where $\mathcal{L}_{\rm max}$ is the maximum likelihood. All the best-fit $\chi^2$'s have been obtained with COSMOMC. In order to have a better statistical comparison taking into account the number of parameters of each model, we also give the differences in the Akaike Information Criterion \cite{AIC} $\Delta$AIC (taking $\Lambda$CDM as base model) defined as AIC$=-2 {\rm ln}(\mathcal{L}_{\rm max})+2k$, with $k$ the number of parameters.} 
\label{tab:DRdatasets}
\end{table}

\subsection{Including \texorpdfstring{$S_8$}{} measurements}

The constraints derived from the CMB+BAO+SnIa data in the preceding section show a lack of discriminating power to {\it measure} the coupling constants $b$ and $\alpha$ and only upper bounds on the strength of the couplings can be obtained. This is not particularly surprising and, in fact, were the interactions actually absent in our Universe, upper bounds on $b$ and $\alpha$ would be the best we could do. However, since we have seen that weaker clustering can be achieved with a stronger interaction without affecting other cosmological parameters, a natural question would be if including additional data with preferred lower values of $\sigma_8$ might favour to remain at higher values of the coupling constants and, ultimately, the non-interacting scenario could be excluded. In this respect, it is interesting to recall that the $\sigma_8$ tension corresponds to a trend from low-redshift probes that measure lower values of $\sigma_8$ as compared to the Planck 2018 for $\Lambda$CDM. Thus, one might expect that adding data from these low-redshift probes could substantially improve the information on the interaction parameters as we have discussed. This expectation is strengthen in view of the results in similar scenarios 
(e.g., \cite{Pourtsidou:2016ico,Jimenez:2020ysu,Figueruelo:2021elm}),
although it is not completely obvious and it will of course depend on the constraining power of the additional data. For instance, the analysis performed in Ref.~\cite{Kumar:2017bpv} for a model with an elastic Thomson-like scattering between DE and DM introduced 
in Ref.~\cite{Simpson:2010vh} shows that the inclusion of the full likelihood of the weak lensing data from CFHTLenS \cite{Heymans:2013fya} does not improve the constraining power of CMB and BAO data 
regarding the interaction parameter. 
On the other hand, it has been found that the inclusion of cluster counts data crucially improves the constraints on the interaction parameters of similar scenarios in Ref.~\cite{Pourtsidou:2016ico} and the forecast analysis carried out in Ref.~\cite{Figueruelo:2021elm} for the $\alpha$CDM model (without varying $N_{\rm eff}$) demonstrates that future surveys like J-PAS, Euclid or DESI will have sufficient discriminating power to detect a non-vanishing interaction parameter at $\sim 5\sigma$. 

In this work, we will not enter into a detailed confrontation of the elastic interacting models with the full data of galaxy surveys for cosmic shear, galaxy-galaxy lensing and galaxy clustering, neither independently nor in their combination commonly called $3\times2$pt. Instead, we will employ an admittedly oversimplified method to study the effect of adding the low-redshift data that suggests smaller values of $\sigma_8$. These data show a certain degeneracy direction in the $\sigma_8$ - $\Omega_m$ plane, so it is convenient to work with the combination\footnote{It is customary in the literature to use $\alpha$ to identify the degeneracy direction of weak lensing 
in the $\sigma_8$ - $\Omega_m$. Since we have already used $\alpha$ for the interaction parameter of the $\alpha$CDM model, we will instead use $p$.}
\be
S_{8,p}\equiv\sigma_8\left(\frac{\Omega_m}{\Omega_{m,\rm fid}}\right)^p\,,
\ee
where different values of $p$ and $\Omega_{m,\rm fid}$ 
are chosen in the literature. Thus, our oversimplified method consists in adding the following Gaussian likelihood: 
\be
\log\mathcal{L}_{S_{8}}=-\frac{(S_{8,\rm model}-S_{8,\rm obs})^2}{2\sigma_{S_8}^2}
\label{eq:GaussianS8}
\ee
with $S_{8,\rm obs}$ the value provided by the corresponding probe and $\sigma_{S_8}$ its corresponding error. A sum over all included values from different surveys is understood. Obviously, we are simply adding a Gaussian prior on $S_8$ as inferred from some external experiment. Remarkably, we will show how adding these data allows to {\it measure} the interaction parameters and this is achieved without worsening any tensions because the resulting contours in the $\sigma_8$ - $\Omega_m$ plane are shifted towards the regions preferred by cluster counts and weak lensing data.

Before delving into the proper analysis and presenting our results, it is convenient to say a few words about possible caveats and shortcomings of our analysis. Firstly, the reported values of $S_8$ correspond to analysis performed for the standard $\Lambda$CDM model. While one might expect a mild dependency of the value of $S_8$ on the fiducial model that has been used to analyse the data, this assumption must eventually be tested on a case by case procedure. On the other hand, the data for cluster counts are usually obtained from a modelling of small scales physics (including some non-linear effects) that uses $\Lambda$CDM as a fiducial model. Given that the couplings we are considering in this work for the DM component provide a substantial modification to the clustering already at the lineal level, it is unclear how much the non-linear clustering will be affected. Thus, an assumption we have to make is that the mass bias and halo mass function are not substantially modified (at least not in a crucial manner). Notice that this is already an issue by itself even within $\Lambda$CDM. In fact, lowering the mass bias factor could alleviate the $\sigma_8$ tension (see e.g., Ref.~\cite{Ade:2013lmv} or the recent paper \cite{Blanchard:2021dwr} for an interesting re-interpretation of the $\sigma_8$ tension). 
The elastic nature of the interacting models under consideration in this work might be useful in this respect because, by their own definition, only the Euler equation of CDM is modified, while the relation of the CDM distribution to the lensing potential and the velocity fields\footnote{Let us notice that galaxies are treated as virialised objects attached to CDM haloes. Thus, they still work as tracers of the underlying gravitational potential that is related as usual to the density field. The interaction is assumed to be sufficiently weak, so galaxies are not off the CDM haloes.}, together with the background evolution, are the same as in $\Lambda$CDM (or $w$CDM in the case of $b$CDM). Thus, the differences are expected to arise only from the different clustering evolution and this might prevent a strong variation in the inferred values of $S_8$ for $\Lambda$CDM and the interacting models. Some of the discussed issues can only be tackled by running dedicated $N$-body simulations that include the interactions under consideration, while other could be resolved by considering the full likelihood data of the corresponding probes. We leave these questions for future work and will focus here on our simplified method to put forward how these data (or, more precisely, how a model-independent measurement of $S_8$) crucially affect the possibility of determining the interaction parameters. 

\begin{figure}
\includegraphics[width=0.48\textwidth]{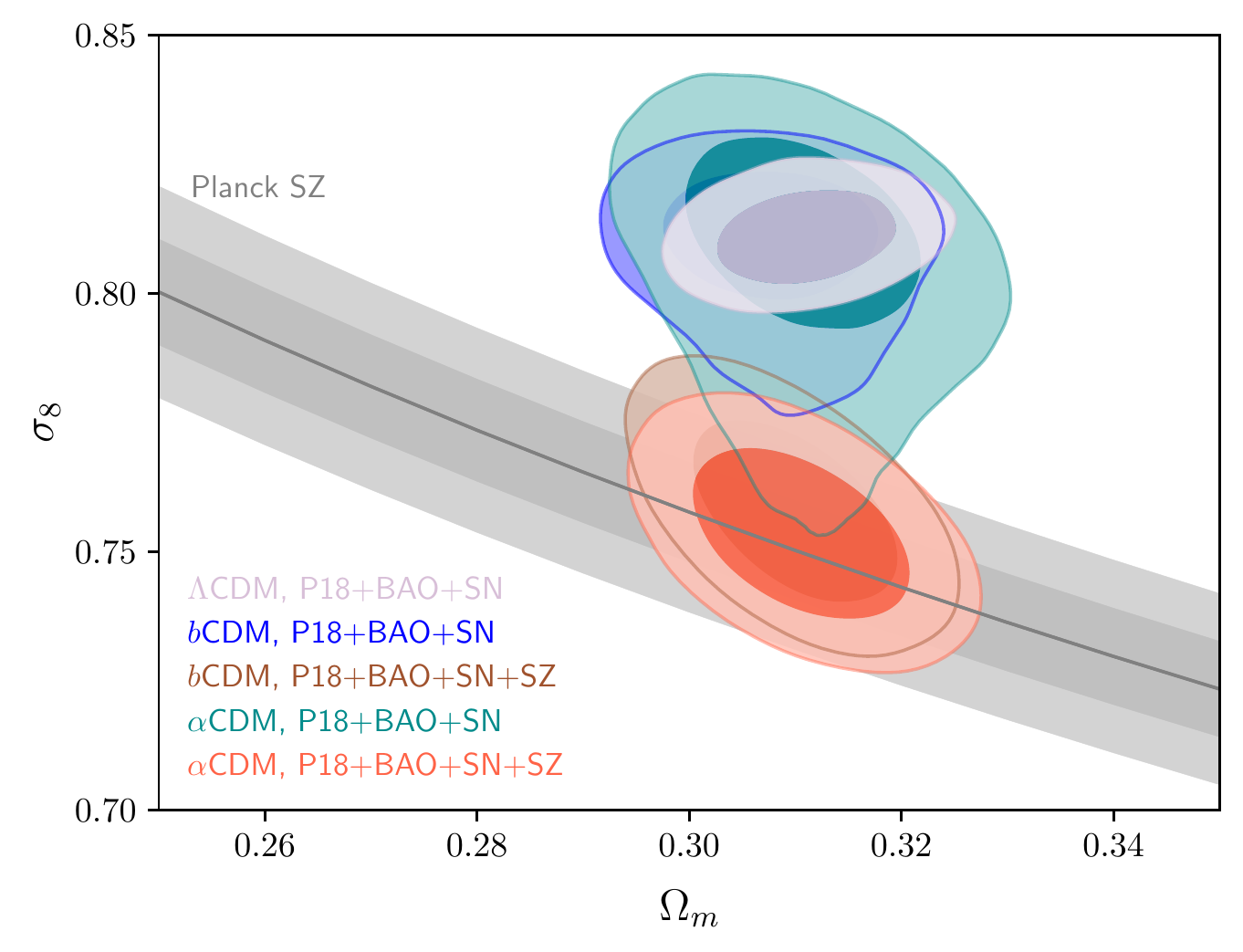}
\hspace{0.5cm}
\includegraphics[width=0.48\textwidth]{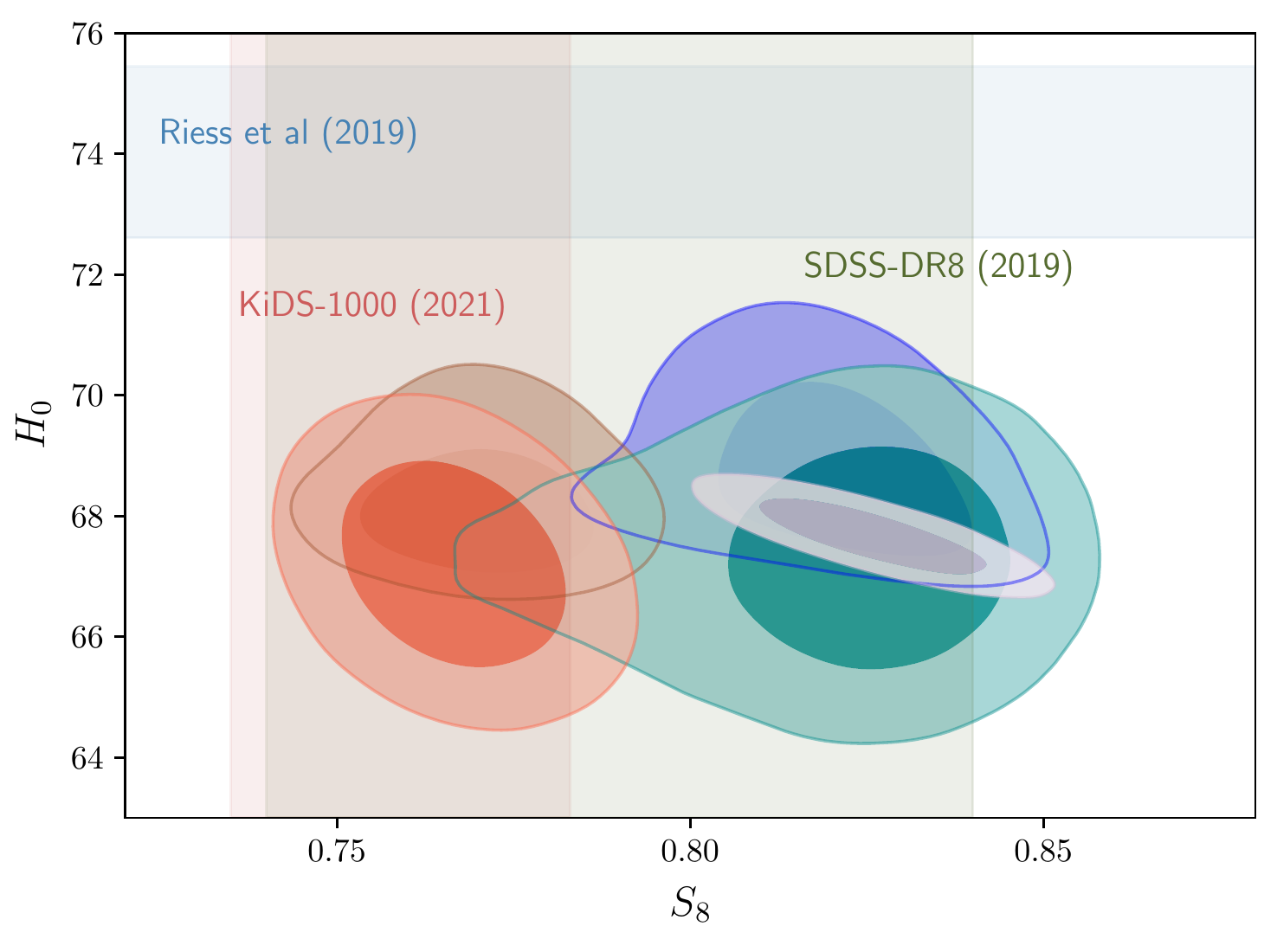}
\caption{Comparison among observational constraints for 
the $\Lambda$CDM, $b$CDM, and $\alpha$CDM models with several data sets. 
{\bf Left}. Limits on the $\Omega_m$ - $\sigma_8$ plane. The grey bands correspond to the 
1-$\sigma$ and 2-$\sigma$ Planck Sunyaev–Zeldovich clusters counts (SZ) $\sigma_8(\Omega_m/0.27)^{0.3} = 0.782 \pm 0.01$ \cite{Ade:2013lmv}. {\bf Right}. Constraints on the $S_8$ - $H_0$ plane, with $S_8=\sigma_8\left(\Omega_m/0.3\right)^{0.5}$. The vertical bands correspond to representative measurements on $S_8$ coming from weak lensing (KiDS-1000 \cite{Asgari:2020wuj}) and cluster counts (SDSS-DR8 \cite{Costanzi:2018xql}), respectively, while the horizontal blue band indicates the 1-$\sigma$ limit for the $H_0$ measurement obtained by theSH0ES collaboration \cite{Riess:2019cxk}. 
}
\label{Fig:omsig8ANDs8H0}
\end{figure}

After explaining our approach and discussing its limitations, we will proceed to the corresponding analysis. We will use the value of $S_8$ obtained from Planck Sunyaev-Zeldovich cluster counts \cite{Ade:2013lmv}
\be
S_{8,{\rm SZ}} \equiv\sigma_8 
\left( \Omega_m/0.27 \right)^{0.3}=0.782 \pm 0.010\,.
\label{eq:S8SZ}
\ee
This value is found in Table 2 of Ref.~\cite{Ade:2013lmv} as obtained from the combination of Planck2013+BAO+BBN and for a fixed mass biased $1-b=0.8$. It is rarely quoted in the literature so our choice might seem arbitrary and it is. However, we choose this single value to have a direct comparison with the 
results in Ref.~\cite{Pourtsidou:2016ico} where the same simplified method was used with the above value of $S_8$ and for a similar scenario. Let us recall again that we do not intend to obtain precise constraints from this data on our interacting scenarios, but rather to illustrate as a proof-of-concept how this information is able to rule out the non-interacting situation. In the following we will refer to the inclusion of the Gaussian likelihood \eqref{eq:GaussianS8} with the value \eqref{eq:S8SZ} as Planck SZ data, bearing in mind our previous discussion on our method. It is convenient to make a further comment regarding the combination of this additional data with those of the preceding section. Since the measurements of $S_8$ from local probes are in tension with the value obtained from Planck, it might be inappropriate to combine both. 
Within our interacting scenarios, the tension is significantly reduced (see Fig.~\ref{Fig:omsig8ANDs8H0}), so it is more consistent to combine the measurement of $S_8$ with Planck data. It is important to keep in mind that Fig.~\ref{Fig:omsig8ANDs8H0} corresponds to marginalised 
2-dimensional distributions and this could hide existing tensions in the data (see e.g., Ref.~\cite{Lin:2017ikq}). 
The absence of very strong correlations in the 
2-dimensional posteriors or multi-modal distributions together with the forecast analysis performed 
in Ref.~\cite{Figueruelo:2021elm} for $\alpha$CDM (without varying $N_{\rm eff}$) make us confident on the reliability of our results, although a more exhaustive analysis would be necessary to rule out said tensions in a robust way.

The effects of including the $S_8$ data for $b$CDM and $\alpha$CDM can be seen in Figs.~\ref{Fig:triangleDR} and \ref{Fig:FitElastic_triangle}. As it is apparent from the $\sigma_8$ - $\log_{10}|b|$ and $\sigma_8$ - $\log_{10}\alpha$ planes, the lower bounds on the interaction parameters disappear for both models and a clear {\it detection} of the interactions is obtained, with values
\ba
\log_{10} \vert b\vert&=&{-1.3}_{-0.4}^{+ 0.2} {}_{-0.6}^{+1.2}\,,\qquad\;\;\;\; (b{\rm CDM})\,,\\
\log_{10} \alpha&=&-0.05_{-0.10-0.28}^{+0.16+0.26}\,,
\qquad (\alpha{\rm CDM})\,.
\ea
Moreover, the detection of the interaction occurs in the region of lower values of $\sigma_8$, so the contours in the $\sigma_8$ - $\Omega_m$ shift vertically towards smaller values of $\sigma_8$ when including the $S_8$ data. As we show in Table \ref{tab:DRdatasets}, the joint analysis using the CMB+BAO+SnIa+SZ data gives the following bounds
\ba
& &
\sigma_8=0.758^{+0.011+0.024}_{-0.012-0.022}\,,\qquad 
\Omega_m=0.31^{+0.006+0.012}_{-0.006-0.013}\,,\qquad \;\;(b{\rm CDM})\,,\\
& &
\sigma_8=0.753^{+0.011+0.022}_{-0.011-0.021}\,,\qquad 
\Omega_m=0.311^{+0.007+0.014}_{-0.007-0.013}\,,\qquad (\alpha{\rm CDM})\,.
\ea
In both models, the constrained values of $\sigma_8$ are smaller 
than those in $\Lambda$CDM (signalling the alleviation of the $\sigma_8$ tension), while the values of $\Omega_m$ are similar to those in $\Lambda$CDM. This property can also be clearly seen in Fig.~\ref{Fig:omsig8ANDs8H0}, which shows the shifts of $\sigma_8$ in the interacting models when adding the $S_8$ information. 
This can be easily understood because the DE-CDM momentum exchange tends to suppress the growth of structures without affecting the abundance of matter. This is in the right direction because the inclusion of galaxy clustering to break the degeneracy between $\sigma_8$ and $\Omega_m$ seems to signal that the tension is driven by $\sigma_8$ rather than $\Omega_m$. Thus, the interacting models under consideration would seem to provide a natural alleviation of the tension. This same behaviour was also found in Ref.~\cite{Pourtsidou:2016ico} for an analogous but different model. In fact, the effect of the momentum exchange in the $\sigma_8$ - $\Omega_m$ plane found in Ref.~\cite{Pourtsidou:2016ico} very closely resembles the one also found here.

The relevant planes for the $H_0$ and $\sigma_8$ tensions are shown in Fig.~\ref{Fig:omsig8ANDs8H0}. In the left panel we see how the interacting models with Planck2018+BAO+SN produce $2\sigma$ contours that suitably spread towards the band of the PlanckSZ cluster counts measurement (grey band), thus improving the compatibility with this data as compared to the $\Lambda$CDM model. The right panel shows the 
$S_8$ - $H_0$ plane where we corroborate how Planck2018+BAO+SN is compatible with lower values of $S_8$ (which directly translates into lower values of $\sigma_8$ because $\Omega_m$ is not substantially affected). Furthermore, not only this improvement in the value of $S_8$ does not occur at the expense of worsening the $H_0$ tension, but the presence of extra radiation actually allows for higher values of $H_0$ than $\Lambda$CDM. This in turn constitutes a better agreement with the SH0ES measurement (horizontal band). The constraints on the 
Hubble parameter are
\ba
& &
H_0=68.14^{+0.45+1.67}_{-0.86-1.32}\;\Hunits,\qquad (b{\rm CDM})\,,\label{H0bf}\\
& &
H_0=67.20^{+1.12+2.24}_{-1.12-2.19}\;\Hunits,
\qquad (\alpha{\rm CDM})\,.
\label{H0alf}
\ea
Although the mean values for the interacting scenarios are similar to those of $\Lambda$CDM, the $2\sigma$ bounds are considerably broader, by a factor of $2 \sim 3$ (as we found in the previous section as well), thus ameliorating the compatibility with the SH0ES values. This is of course at the expense of adding additional parameters, so one would need to check that paying the price is worth it. 

\begin{figure}[ht!]
\includegraphics[width=1\textwidth]{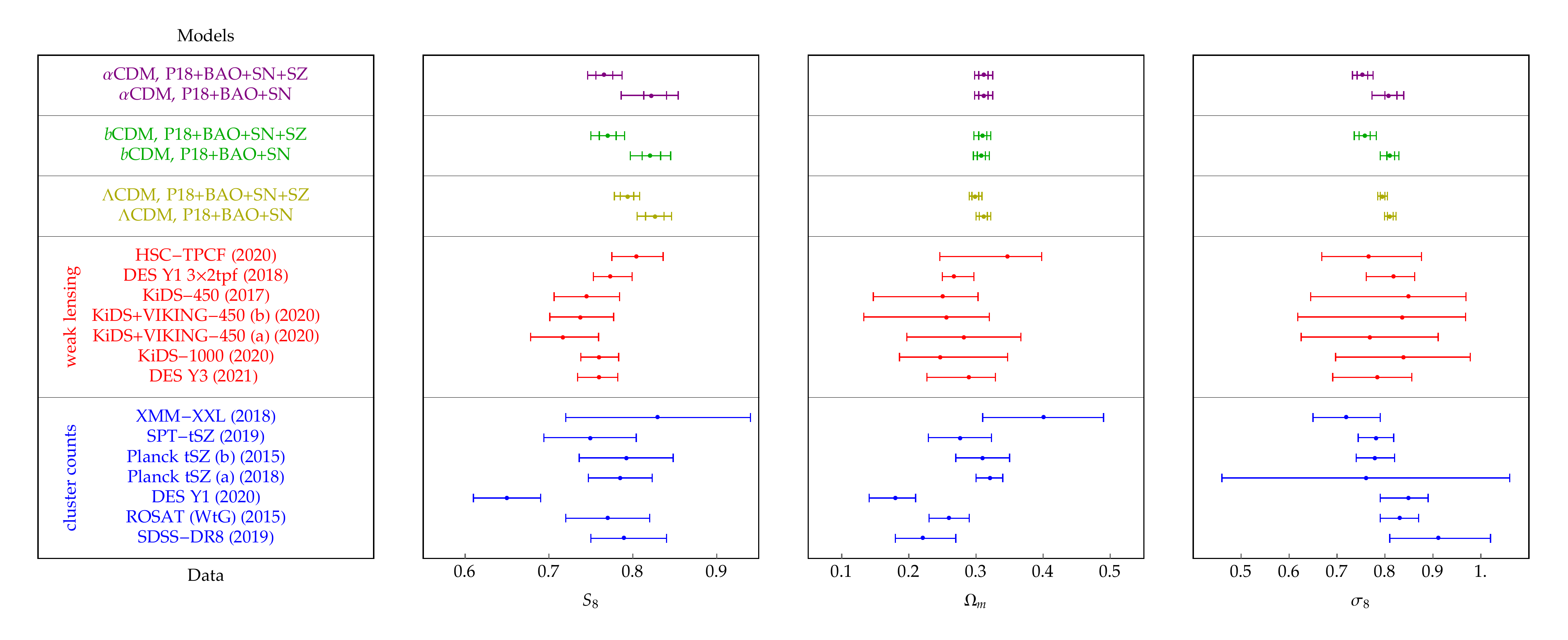}
\caption{We compare the values obtained for $S_8$, 
$\Omega_m$, and $\sigma_8$ for the interacting scenarios and to those inferred for $\Lambda$CDM from several weak lensing and cluster counts probes. 
Cluster counts data: SDSS-DR8 (2019) \cite{Costanzi:2018xql}, ROSAT (WtG) (2015) \cite{Mantz:2014paa}, DES Y1 (2020) \cite{Abbott:2020knk}, Planck tSZ (a) (2018) \cite{Ade:2015fva}, Planck tSZ (b) (2015) \cite{Salvati:2017rsn}, SPT-tSZ (2019) \cite{Bocquet:2018ukq}, XMM-XXL (2018) \cite{Pacaud:2018zsh}; weak lensing data: 
DES Y3 (2021) \cite{Amon:2021kas},
KiDS-1000 (2020) \cite{Asgari:2020wuj},  
KiDS+VIKING-450 (a) (2020)  \cite{Wright:2020ppw},
KiDS+VIKING-450 (b) (2020) \cite{Hildebrandt:2018yau}, 
KiDS-450 (2017) \cite{Hildebrandt:2016iqg}, DES Y1 3x2tpf (2018) \cite{Abbott:2017wau}, 
HSC-TPCF (2020) \cite{Hamana:2019etx}.
}
\label{Fig:s8dataDR}
\end{figure}

\begin{figure}[ht!]
\includegraphics[width=0.8\textwidth]{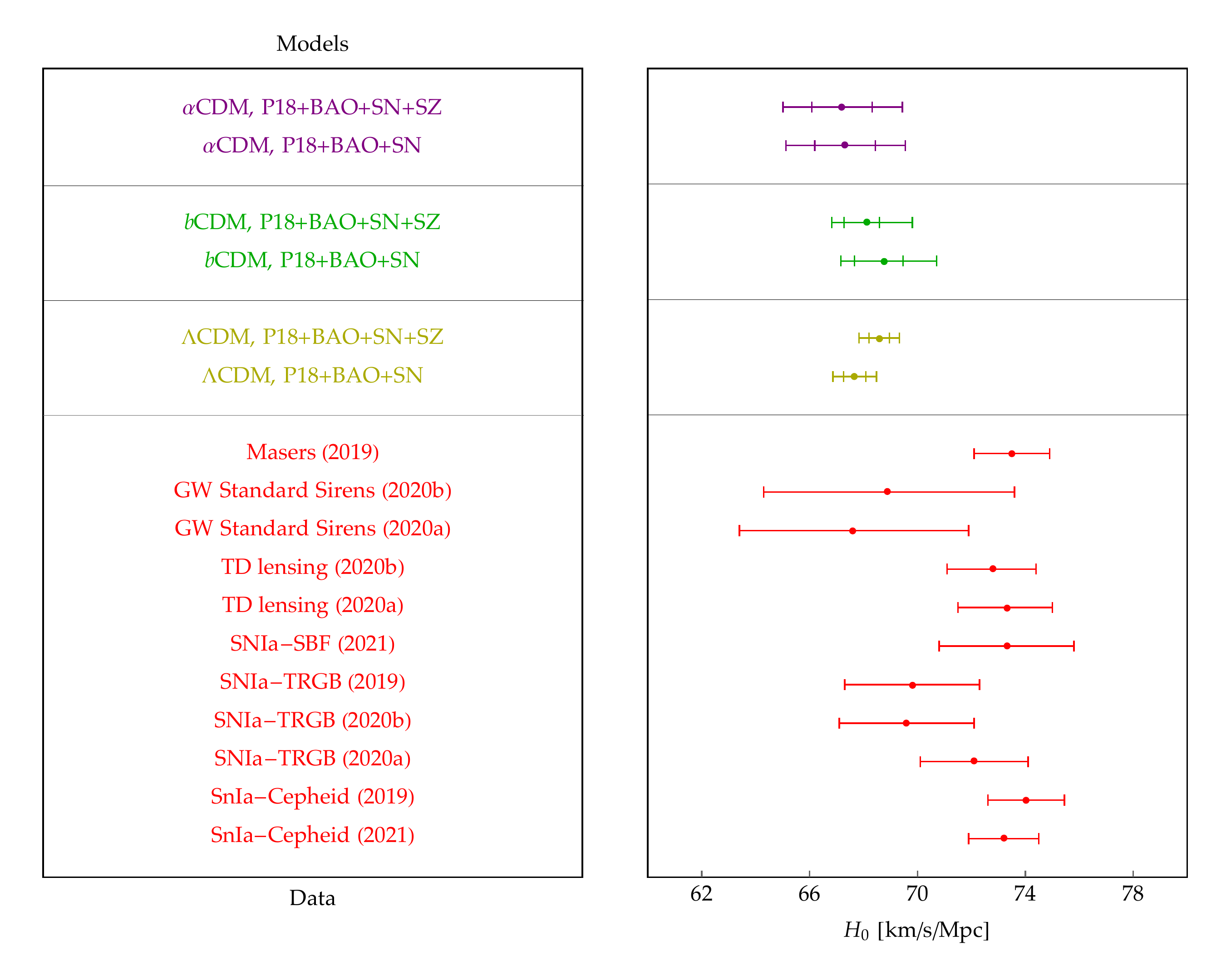}
\caption{In this whisker plot we compare the constraints on $H_0$ obtained in this paper for $\Lambda$CDM, $b$CDM and $\alpha$CDM to some local measurements obtained by different methods: SnIa-Cepheid (2021) \cite{Riess:2020fzl}, SnIa-Cepheid (2019) \cite{Riess:2019cxk}; SNIa-TRGB (2020a) \cite{Soltis:2020gpl}; SNIa-TRGB (2020b) \cite{Freedman:2020dne}; SNIa-TRGB (2019) \cite{Freedman:2019jwv}; SNIa-SBF (2021) \cite{Blakeslee:2021rqi}; TD lensing (2020a) \cite{Wong:2019kwg}, TD lensing (2020b) \cite{Liao:2020zko}; GW Standard Sirens (2020a) \cite{Mukherjee:2020kki}, GW Standard Sirens (2020b) \cite{ Hotokezaka:2018dfi}; Masers (2019) \cite{Reid:2019tiq}.} 
\label{Fig:dataH0}
\end{figure}

In Table \ref{tab:DRdatasets} we also provide the best-fit values of $\chi^2$ for the different models. Without the information on $S_8$, the three models show similar performances and the AIC criterion does not reveal any statistical preference for any of the interacting models. However, when including the information on $S_8$, the $\chi^2$ values for the interacting models are drastically reduced with respect to $\Lambda$CDM and they are strongly favoured even with the penalty of having more parameters according to the AIC criterion. Clearly, the PlanckSZ data drives this significant improvement and this is caused by the presence of elastic interactions that permit to have lower values of $\sigma_8$ without altering $\Omega_m$. 
Finally, Fig.~\ref{Fig:s8dataDR} shows a compilation of the values of $S_8$ obtained by weak lensing and cluster counts measurements reported by different surveys within the $\Lambda$CDM framework. We also show the values obtained for our interacting scenarios. In principle, our analysis with the PlanckSZ cluster counts value could be extended to include the values of other probes, barring the discussed caveats of our method. However, we do not expect qualitative changes in the results, although quantitative differences will be obtained. 
In Fig.~\ref{Fig:dataH0} we compare the constraints on $H_0$ with some local measurements, where we can see how the $2\sigma$ bounds in the interacting models are wider than in $\Lambda$CDM, thus allowing a certain alleviation of 
the tension.

\section{Discussion}

In this work, we have studied observational 
constraints on a class of models featuring a dark sector with interactions driven by the relative motion of DE and DM. The common property of these models is the presence of a coupling of DM with a pressurefull fluid that reduces the growth of structures at late time. This favours having a weaker clustering and, thus, potentially alleviating the $\sigma_8$ tension. In the two considered scenarios (that we have dubbed $b$CDM and $\alpha$CDM), we have included an additional radiation component to explore its effects on the Hubble tension as well. In $b$CDM the dark radiation is crucial for suppressing the growth of structures because it also represents the fluid with which DM interacts, whereas in $\alpha$CDM the extra radiation via a free $N_{\rm eff}$ does not play a relevant role for lowering the value of $\sigma_8$.

After briefly introducing the models and discussing their main properties, we have confronted their predictions against a relatively standard set of cosmological datasets including temperature, polarisation and lensing data from Planck2018, BAO and SnIa. This analysis has confirmed the initial expectations that the extra radiation component can alleviate the Hubble tension, while the additional interaction works in favour of reducing the $\sigma_8$ tension. However, these models seem to prefer to ease only one tension at a time and the sweet spot in the $H_0$ - $\sigma_8$ plane where both tensions would be simultaneously alleviated is outside the 2$\sigma$ region. 
It is however interesting that improving one tension does not worsen the other as it occurs in other scenarios because the usual (anti-)correlation in the $S_8$ - $H_0$ plane is substantially reduced.

An outcome of our analysis is that the interaction parameter in both cases cannot be constrained by the datasets of Planck2018, BAO and SnIa, but only an upper bound can be set. We have then proceeded to include low redshift data of $S_8$ that provide constraints in the $\Omega_m$ - $\sigma_8$ plane. We have shown how the interaction in the dark sector improves the compatibility between Planck data and the values of $S_8$ reported by different analysis. These comparisons have to be made bearing in mind that the measurements are obtained by assuming a $\Lambda$CDM fiducial model to deal with the raw data (calibrations, mass bias, hydrodynamical modelling of the gas, etc.) and the dependence on such a fiducial model varies depending on the particular analysis. Nevertheless, we have resorted to a simplified approach where we assume a certain model-independence for the reported values of the parameter $S_8$ and included this value as an additional Gaussian likelihood to illustrate how the inclusion of these data allows the determination of 
the interaction parameters. We have shown how the 1-dimensional marginalised posterior of the interaction parameters in both models peaks at a certain value with the non-interacting case at a few $\sigma$s, i.e., the inclusion can potentially signal to the detection of the couplings. This seems to be a generic property of this type of models because a similar result was obtained in Ref.~\cite{Pourtsidou:2013nha}, where a pure-momentum exchange with a scalar field was considered. 
Of course, one needs to be careful before claiming the detection of an interaction in the dark sector and more robust analysis would be necessary to strengthen the significance of this result. For instance, it is known that marginalised posterior distributions can mask the existence of tensions between different datasets and this could lead to a fake detection of cosmological parameters such as a neutrino mass or an interaction in the dark sector as in our case. In our case, we have seen that the interacting models can substantially reduce the tension between Planck 2018 and the low-redshift measurements of $S_8$. Since this is precisely the crucial data to obtain a detection of the interaction parameters, we should analyse more exhaustively the compatibility of the employed data within the interacting scenarios to unveil potential inconsistencies in the entire posterior distributions.

Our results in this work are in agreement with those already obtained in the literature for similar scenarios. It is remarkable how these models consistently alleviate the cosmological tensions and, furthermore, some low redshift datasets related to clustering (weak lensing, cluster counts) seem to strongly prefer them over the standard $\Lambda$CDM model, barring the discussed caveats regarding the use of these data. The obtained promising results motivate further studies in order to clarify the true viability of these models. An important question is of course the already discussed issue of clarifying how the non-linear structure formation can affect the picture derived by using the linear evolution only. 
Simulations for elastic interactions in the dark sector are scarce in the literature, so it is unclear what to expect from non-linear effects on small scales. An elastic Thomson scattering between DE and DM as discussed 
in Ref. \cite{Simpson:2010vh} was simulated in Refs.~\cite{Baldi:2014ica,Baldi:2016zom}. It is not clear how much intuition one can gain from their results for our scenarios given the different eras at which the interactions occur, but their results in Ref.~\cite{Baldi:2016zom} confirm the potential of these models to alleviate the $\sigma_8$ tension when including non-linear effects.

Even without resorting to non-linear evolution, a more thoroughly dedicated analysis of weak lensing and cluster counts data beyond our simplified treatment using $S_{8}$ would be worth pursuing. Let us recall that these constraints are obtained for $\Lambda$CDM model, so using them at face value might not be well-justified were they strongly dependent on the fiducial model. In this respect, let us notice that the analysis of DES-Y3 cosmic shear analysis for $w$CDM gives slightly smaller values for $S_8$ \cite{Secco:2021vhm} than for $\Lambda$CDM model although both results are well within $1\sigma$. This might suggest that, as a first approach to get a rough estimate, using $S_8$ as we have done in this work could suffice, but it is clear that a more proper analysis is required to draw a more robust conclusion. However, a compelling feature of our interacting scenarios is their minimalism since the background evolution, the relation of the gravitational potentials to the density field and the relation of the density to the velocity field of CDM are the standard ones. The only modifications arise for the equations of 
velocity potentials 
and a deviation between the baryon and 
DM evolutions that could affect the bias, i.e., how faithfully galaxies trace the underlying density field. 
These properties could be relevant for reducing potential differences in the measurement of $S_8$ when using $\Lambda$CDM and the interacting models.

Another useful source of information to constrain this class of models will be the growth rate data, since the main distinctive property of these models is precisely the significantly different evolution of the growth rate as compared to $\Lambda$CDM model, specially at low redshift. However, one needs to be careful to properly account for the scale-dependence of $\delta_{c}$ when comparing to data. Finally, the full shape of the power spectrum will also be helpful to constrain these interactions. These are future directions that are currently under way and whose results will be reported elsewhere.  \\

{\bf Codes:} The modified versions of the codes CAMB and CLASS using for this work  are available upon request.


{\bf Acknowledgements:} We thank Osamu Seto for useful discussions.
JBJ, DB, DF and FATP acknowledge support from the {\it Atracci\'on del Talento Cient\'ifico en Salamanca} programme, from project PGC2018-096038-B-I00 by {\it  Spanish Ministerio de Ciencia, Innovaci\'on y Universidades} and {\it Ayudas del Programa XIII} by USAL. DF acknowledges support from the programme {\it Ayudas para Financiar la Contrataci\'on Predoctoral de Personal Investigador (ORDEN EDU/601/2020)} funded by Junta de Castilla y Le\'on and European Social Fund. DB acknowledges support from Junta de Castilla y Le\'on and Fondo Europeo de Desarrollo Regional (FEDER) through the project SA0096P20. ST is supported by the Grant-in-Aid for Scientific Research Fund of the JSPS No.\,19K03854. \\

\bibliography{bibinteractingfluids}

\end{document}